\documentclass[twocolumn,groupedaddress,aps,pra,amsmath,amssymb,showpacs,10pt]{revtex4-2}
\usepackage[colorlinks,linkcolor=blue,anchorcolor=blue,citecolor=blue,bookmarksnumbered]{hyperref}
\usepackage{graphicx,epsfig,float,epstopdf}
\usepackage{amssymb,amsmath,subeqnarray,mathrsfs,amsfonts,bm,makecell,cases,color,extpfeil,lipsum}
\usepackage{comment}

\allowdisplaybreaks

\usepackage{graphicx}
\usepackage{dcolumn}
\usepackage{bm}

\begin{document}

\title{Shock wave generation and propagation in dissipative and nonlocal nonlinear Rydberg media}

\author{Lu Qin$^{1,2,3}$, Chao Hang$^{4}$, Guoxiang Huang$^{4}$, and Weibin Li$^{2,3,}$}
\affiliation{
$^{1}${Department of Physics, Henan Normal University, Xinxiang 453007, China}\\
$^{2}${School of Physics and Astronomy, University of Nottingham, Nottingham, NG7 2RD, United Kingdom}\\
$^{3}${Centre for the Mathematics and Theoretical Physics of Quantum Non-Equilibrium Systems, University of Nottingham, Nottingham NG7 2RD, United Kingdom}\\
$^{4}${State Key Laboratory of Precision Spectroscopy, East China Normal University, Shanghai 200062, China}
}

\date{\today}

\begin{abstract}
We investigate the generation of optical shock waves in strongly interacting Rydberg atomic gases with a spatially homogeneous dissipative potential. The Rydberg atom interaction induces an optical nonlocal nonlinarity. We focus on local nonlinear ($R_b\ll R_0$) and nonlocal nonlinear ($R_b\sim R_0$) regimes, where $R_b$ and $R_0$ are the characteristic length of the Rydberg nonlinearity and beam width, respectively. In the local regime, we show spatial width and contrast of the shock wave change monotonically when increasing strength of the dissipative potential and optical intensity. In the nonlocal regime,  the characteristic quantity of the shock wave depend on $R_b/R_0$  and dissipative potential nontrivially and on the intensity monotonically. We find that formation of shock waves dominantly takes place when $R_b$ is smaller than $R_0$, while the propagation dynamics is largely linear when $R_b$ is comparable to or larger than $R_0$. Our results reveal nontrivial roles played by dissipation and nonlocality in the generation of shock waves, and provide a route to manipulate their profiles and stability. Our study furthermore opens new avenues to explore non-Hermitian physics, and nonlinear wave generation and propagation by controlling dissipation and nonlocality in the Rydberg media.
\end{abstract}
\maketitle

\section{Introduction}
Nonlinear hydrodynamic flows are found in different media~\cite{whitham1999linear,ablowitz2011nonlinear}, such as plasma~\cite{TaylorPRL1970,romagnani2008PRL,pakzad2011ASS}, acoustics systems~\cite{gurbatov2012book}, polymerized ionic liquid~\cite{LeeACS2019}, quantum-mechanical piston~\cite{mossman2018NC,hoefer2008PRL}, ultracold quantum gases~\cite{PerezPRL2004,kamchatnov2004PRA,hoefer2006PRA,joseph2011PRL,chang2008PRL,Simmons2023PRA}, and optical media~\cite{el2007PRA,rothenberg1989PRL,wan2007NP,jia2007PRL,ghofraniha2007PRL,Kartashov2013OL,gentilini2015PRA,marcucci2019APX,xu2017PRL,WetzelPRL2016,ContiPRA2010,marcucci2020aPRL}. In these systems, the competition of nonlinearity, dispersion and dissipation gives rise to nonlinear wave phenomena,
such as solitons~\cite{bai2019Optica,chen2014PRA,hang2018PRA}, and rogue wave~\cite{liu2016PRA,liu2017weak}. In defocusing nonlinear media, an initially smooth wave steepens when propagation, eventually reaching a point of gradient catastrophe~\cite{kamchatnov2000book},  that lead
to the formation of shock waves~\cite{courant1944book,kamchatnov2000book,kleine2001book,zel2002book,ghofraniha2012PRL,Hang2023PRA,mossman2018NC,isoard2019PRA,isoard2020EPL,Kamchatnov2020chaos,el2016PRS,Simmons2020PRL,smoller2012book,marcucci2019NC,bienaime2021PRL}. Profiles of shock waves depend on the dissipation of the medium. In dissipation free media, the formed dispersive shock waves (DSW) show a strong oscillatory structure due to the interplay between the nonlinearity and dispersion~\cite{el2007PRA,wan2007NP,lowman2013PRA,Hang2023PRA,GarnierPRL2013,ArmaroliPRA2009,El2017SIAM,Maiden2016PRL,Kartashov2012OL}. This steepening can also be mediated by dissipation, where the nonlinear wave acquires a monotonic shock
front without any oscillations. In such cases, a dissipative shock wave, sometimes also called viscous shock wave (VSW), 
emerges~\cite{HerboldPRE2007,mossman2018NC,lowman2013PRA,LeeACS2019}.

Recently, it has been shown that cold atomic gases interacting with laser fields provide a fertile ground for studying shock waves~\cite{PerezPRL2004,kamchatnov2004PRA,hoefer2006PRA,joseph2011PRL,chang2008PRL,Simmons2023PRA,kamchatnov2004PRA}. 
When additionally coupling the light to highly excited Rydberg states~\cite{saffman_quantum_2010,shao_2024},  strong and long-range interactions between Rydberg atoms can be mapped to light fields through electromagnetically induced transparency (EIT)~\cite{PhysRevLett.107.133602}, generating strong \textit{nonlocal nonlinearities}~\cite{sevinccli2011PRL,li_electromagnetically_2014}. The characteristic length of the nonlocal nonlinearity, determined by the blockade radius of the Rydberg gas, is in the order of micrometers, which is comparable to typical beam width.
Using the strong Rydberg nonlocal nonlinearity (NNL), it has been shown that DSWs can be generated and manipulated in Rydberg atom gases~\cite{Hang2023PRA}.
Dissipation plays an important roles in the study of Rydberg systems~\cite{yan_electromagnetically_2020,hao_observation_2021,dingErgodicityBreakingRydberg2024}. In cold atom gases, dissipation, on the other hand, can be induced and controlled~\cite{hang2018PRA,bai2019Optica}. This opens new opportunities for exploring shock waves in the interplay between the nonlocal nonlinearity and controllable dissipation that is otherwise difficult to achieve in other systems. 

In this work, we study the generation and propagation of shock waves within a cold Rydberg atomic gas setting, incorporating an engineered, homogeneous dissipative potential that can be changed from loss to gain. This change is controlled by employing an incoherent pumping~\cite{hang2018PRA} and controlling the laser detuning~\cite{bai2019Optica}. A nonlocal optical nonlinear interaction is induced by coupling low-lying electronic states to Rydberg $S$ state via EIT~\cite{PhysRevLett.107.133602,sevinccli2011PRL}. Depending on the blockade radius $R_b$ of the NNL and beam width $R_0$, the system is in a local regime when $R_b\ll R_0$, and nonlocal regime when $R_b\sim R_0$. In the local regime, the nonlinear Schr\"odinger (NLS) equation governing the light propagation is cast into coupled Riemann equations. Formation of shock waves is signified by wave breaking. Excluding dissipation, wave breaking points are obtained analytically through the Riemann equations. Oscillation contrast~\cite{isoard2019PRA, xu2016OL} and spatial width of the shock wave depends on the strength of the dissipation and optical intensity monotonically in the local regime.
In the nonlocal regime, the nonlocal degree of the optical nonlinearity modifies amplitudes and width of shock waves. Importantly properties of the shock wave exhibit complicated dependence on the nonlocality. We show that shock wave generation and propagation is important when $R_b<R_0$. When $R_b$ and $R_0$ are comparable, the medium is effectively linear, where the NNL becomes a homogeneous dispersive potential approximately.

The paper is arranged as follows. 
In Sec.~\ref{Sec2}, we present the physical model that can lead to the dissipative and nonlocal nonlinear potential. The NLS equation that describes the propagation of the probe laser ﬁeld is derived. 
In Sec.~\ref{sec4}, light propagation in the local regime is discussed. 
The impact of the dissipative potential on the oscillation contrast,  width, and shock width is investigated.
In Sec.~\ref{sec5}, we explore the influence of the nonlocality on the generation and propagation of shock waves. Finally, conclusions are given in Sec.~\ref{sec6}.

\section{Model and light propagation equations}	\label{Sec2}
\subsection{Physical model}

\begin{figure}[t]
\centering
\includegraphics[width=0.92\linewidth]{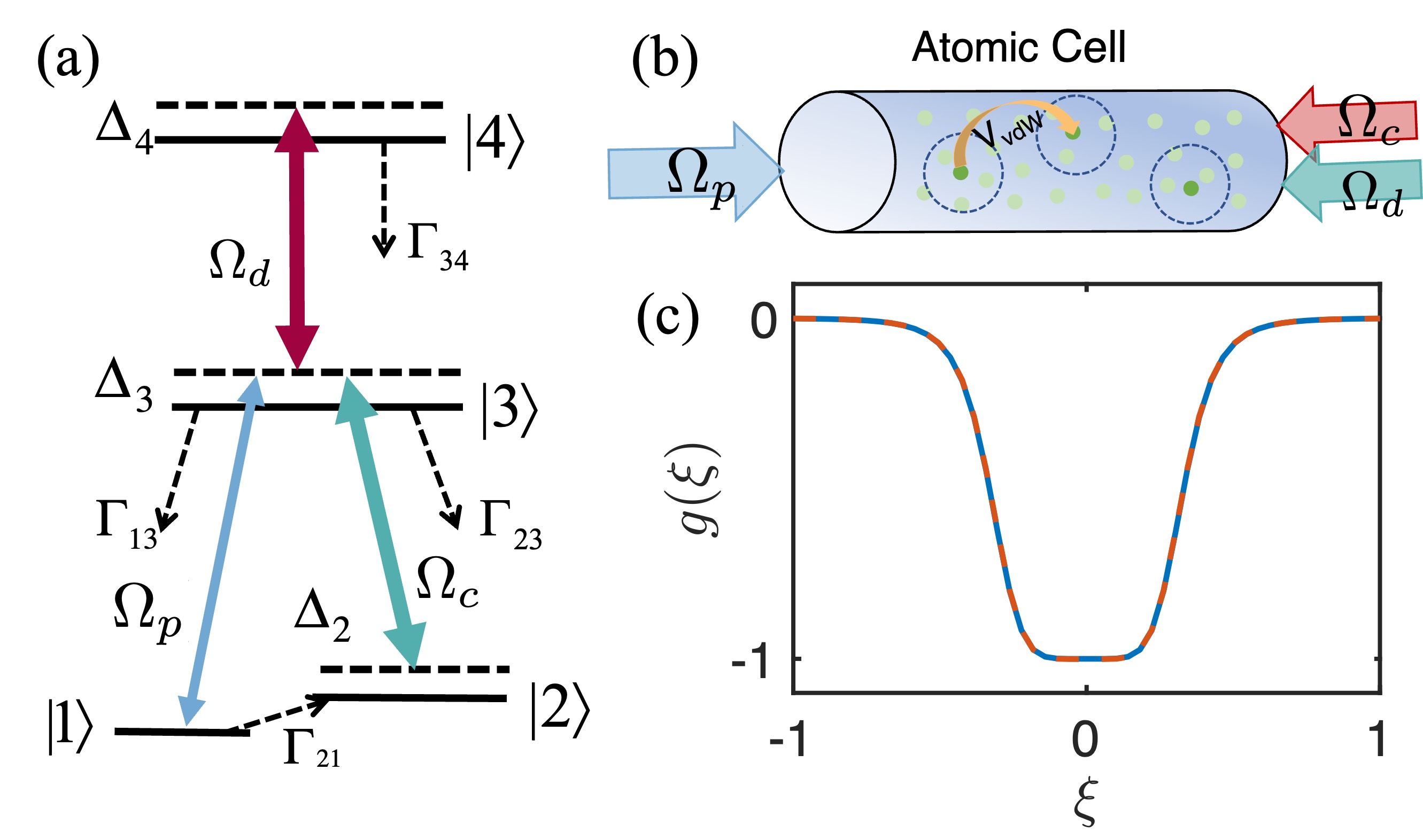}
\caption{{\bf The system}.
(a)~Inverted Y-type level diagram. A weak probe laser field (half Rabi frequency $\Omega_{p}$) couples transition $|1\rangle\leftrightarrow|3\rangle$. A strong control (dressed) laser field [half Rabi frequency ${\Omega}_c$ ($\Omega_d$)] couples transition $|2\rangle\leftrightarrow|3\rangle$ ($|3\rangle\leftrightarrow|4\rangle$). In  Rydberg state $|4\rangle$ atoms interact strongly through the van der Waals interaction. Here $\Delta_{\alpha}$ are detuning and $\Gamma_{\alpha\beta}$ are spontaneous emission decay rates ($\alpha<\beta$). The emission from state $|3\rangle$ generates an effective loss potential. To achieve the gain, we apply an incoherent pumping $\Gamma_{21}$ between states $|1\rangle$ and $|2\rangle$.
(b) The blue, red, and cyan arrows indicate the propagating direction of the probe, control, and dressed fields. The strong Rydberg atom interaction gives rise to an blockade radius (dashed circle) around Rydberg atoms. The blockade radius can be tuned by the excitation laser. (c) Response function $g(\xi)$. The numerically obtained $g(\xi)$ (blue solid line) agrees with the analytical approximation Eq.~(\ref{response function}) (dashed red line). The depth of the potential is given by $1/b_1$. Other parameters are 
$b_1=1$, $b_2=2.6$, 
$B_1=0.001$, and $B_2=0.38$.  See text for details. }
\label{fig_model}
\end{figure}
	
We consider a gas of cold atoms with an inverted Y-type four-level configuration [see Fig.~\ref{fig_model}(a)], where a weak probe laser field with half Rabi frequency $\Omega_{p}$ couples the transitions $|1\rangle\leftrightarrow|3\rangle$. A strong control and a dressed laser fields with half Rabi frequencies ${\Omega}_c$ and $\Omega_{d}$ couple the transition $|2\rangle\leftrightarrow|3\rangle$ and $|3\rangle\leftrightarrow|4\rangle$, correspondingly. Detuning $\Delta_{\alpha}$\,($\alpha=2,\,3,\,4$) gives difference between laser frequency and atomic transition. And $\Gamma_{\alpha\beta}$ are spontaneous emission decay rates from $|\beta\rangle$ to $|\alpha\rangle$ ($\alpha<\beta$). Here the incoherent decay from state $\lvert 3\rangle$ causes loss of the probe field. To generate gain, an incoherent pumping (with pumping rate $\Gamma_{21}$) is used to pump atoms from $|1\rangle$ to $|2\rangle$. Driven by the control laser $\Omega_c$ a small number of atoms are populated in state $|3\rangle$ which  provides a gain effect~\cite{hang2018PRA,hang2019PRA}.

In this setting, state $|4\rangle$ is a high-lying Rydberg state. The interaction between two Rydberg atoms respectively at positions ${\bf r}$ and ${\bf r}^{\prime}$ is described by van der Waals potential $V_{\mathrm{vdW}} \equiv \hbar V(\mathbf{r}^{\prime}-\mathbf{r})= -\hbar C_{6}/|\mathbf{r}'-\mathbf{r}|^{6}$~\cite{gallagher1994book}. 
When the light propagates in the medium [see Fig.~\ref{fig_model}(b)],  Rydberg excitation in the vicinity of a Rydberg atom is strongly suppressed, due to the long-range Rydberg-Rydberg interaction. Such spatial dependent Rydberg blockade leads to nonlocal nonlinear optical interactions~\cite{murray2016quantum}. 
Note that in the inverted Y-shaped excitation scheme shown in Fig.~\ref{fig_model}(a), the transition $|1\rangle \rightarrow|2\rangle \rightarrow|3\rangle$ forms a $\Lambda$-type EIT, while  $|1\rangle \rightarrow|3\rangle \rightarrow|4\rangle$ forms a ladder type  Rydberg-EIT. The interplay of the two paths can be controlled by the external lasers, giving rise to dissipative, nonlocal nonlinear interactions. 

Under electric-dipole and rotating-wave approximations, Hamiltonian
of the system is $\hat{H}={\cal N}_a \int d^3 r\, \hat{\cal H}$ with Hamiltonian density $\hat{\cal H}$ ($\hbar\equiv 1$),
\begin{eqnarray}\label{Hamiltonian}
{\cal \hat{H}}&=-\sum\limits_{\alpha
=1}^{4}{\Delta_{\alpha}\hat{S}_{\alpha \alpha}({\bf r})}- \big[\Omega_{p}\hat{S}_{13}({\bf r})+\Omega_{c}\hat{S}_{23}({\bf r})+\Omega_d\hat{S}_{34}({\bf r}) \notag 
\\
&
+{\rm h.c.} \big]+\mathcal N_{a} \int d^{3} r^{\prime} \hat{S}_{44}({\bf r}^{\prime}) \hbar V ({\bf r}^{\prime}-{\bf r}) \hat{S}_{44} ({\bf r})\nonumber,
\end{eqnarray}
where ${\cal N}_a$ is atomic density, and $\hat{S}_{\alpha\beta}(\mathbf{r})=|\beta\rangle\langle \alpha|\,\exp \{i[(\mathbf{k}_{\beta}-\mathbf{k}_{\alpha})\cdot\mathbf{r}
-(\omega_{\beta}-\omega_{\alpha}+\Delta_{\beta}
-\Delta_{\alpha})t]\}$ is the atomic transition operator between states $|\alpha\rangle$ and $|\beta\rangle$. For weak excitation, $\hat{S}_{\alpha\beta}(\mathbf{r})$ are approximated by bosonic operators~\cite{Lukin2000PRL}.
Taking into account of decay, dynamics of the density matrix (matrix elements $\rho_{\alpha\beta}\equiv\langle\hat{S}_{\alpha\beta}\rangle$) is described by the Bloch equation $\partial \hat{\rho}/\partial t=-i[\hat{H}, \hat{\rho}]/\hbar-\Gamma\,[\hat{\rho}]$,
where $\Gamma$ is the relaxation matrix describing the spontaneous emission and dephasing (see Appendix~\ref{A}). Propagation of the semiclassical probe field is governed by the Maxwell equation,
\begin{eqnarray} 
i\left( \frac{\partial} {\partial z} + \frac{1} {c} \frac{\partial} {\partial t}  \right) \Omega_{p} + \frac{c}{2\omega_{p}}\left(\frac{\partial ^2}{\partial x^2} +\frac{\partial ^2}{\partial y^2}\right)\Omega_{p}
+\kappa_{13}\rho_{31}=0, \nonumber
\end{eqnarray}
where $\kappa_{13}=\mathcal{N}_a\omega_{p}|{\bf p}_{13}|^2/(2\varepsilon_0 c \hbar)$, with $\omega_p$ the weak probe laser field of center frequency, ${\bf p}_{13}$ the electric dipole matrix element associated with the transition $|1\rangle\leftrightarrow|3\rangle$, $\varepsilon_0$ the vacuum dielectric constant, and $c$ the speed of light in vacuum. In deriving the propagation equation, the paraxial and slowly varying envelope approximations have been applied. 

\subsection{Nonlinear envelope equation}

For weak probe fields, the Maxwell and Bloch equations can be solved perturbatively. The Rydberg-Rydberg interaction, on the other hand, is treated beyond the simple mean-field approximation~\cite{bai2016OE,bai2019Optica}. We then solve the Bloch equation up to the third-order of $\Omega_{p}$. This allows to derive a (2+1)D {\it nonlocal nonlinear} Schr\"{o}dinger (NNLS) equation of the probe field~\cite{hang2018PRA,hang2019PRA,shi2023chaos}, 
\begin{eqnarray}\label{NNLS}
i\frac{\partial }{\partial z}\Omega_{p} &+& \frac{c}{2\omega_{p}}\nabla_\bot ^2{\Omega_{p}} -V_1\Omega_p+W{\left| {\Omega_{p}} \right|^2} \\
\qquad&+& \int {d^2{r^{\prime}_{\perp}} G({\bf r}_{\perp},{\bf r^{\prime}_{\perp}}){{\left| {{\Omega_{p}({\bf r^{\prime}_{\perp}},z,t)}} \right|}^2} {\Omega_{p}({\bf r})}}= 0,\nonumber
\end{eqnarray}
with ${\bf r}_{\perp}=(x,y)$. Here $V_1$ is the dissipative potential that gives homogeneous gain or loss in the medium controlled by the laser parameters~\cite{hang2018PRA}. Details on the control of the dissipative potential can be found in Appendix~\ref{B}. 
Nonlinear coefficient $W$ characterizes the local Kerr nonlinearity (contributed by the weak, short-range interactions between photons and atoms~\cite{chen2014PRA}).  $G(\mathbf{r})$ is a nonlocal nonlinear response function characterizing respectively NNL~\cite{bai2016OE,bai2019Optica,qin2020PRA}. 

Compared to the Rydberg induced nonlinearity, the conventional Kerr nonlinearity $W$ is negligible. Using $^{87}$Rb as an example with electronic states $|1\rangle=|5 S_{1/2},\,F=1,\, m_F=-1\rangle$,
$|2\rangle=|5 S_{1/2},\,F=1,\, m_F=1\rangle$,
$|3\rangle=|5 P_{3/2},\,F=1,\, m_F=0\rangle$,
and $|4\rangle=|n S_{1/2}\rangle$, and typical parameters 
$\Delta_2=20$~MHz,
$\Delta_3=-200$~MHz,
$\rm \Delta_4=10$~MHz,
${\rm \Gamma_{3}=2\pi\times6.1}$\,MHz,
$\Gamma_{4}=2\pi\times 16.7$\,kHz, $\Omega_{c}=40$~MHz,
$\Omega_{d}=10$~MHz, and $\mathcal{N}_a=2.3\times 10^{10}\,{\rm cm^{-3}}$, we evaluate the local nonlinearity, characterized by a dimensionless strength $\mathcal{W}=2L_{\rm diff}U_0^2W$. Considering the probe beam radius $R_0=5~\mu$m and $U_0=10$ MHz, we obtain $R_b\approx 1.94 ~\mu{\rm m}$ and $L_{\rm diff}\approx 0.2$~mm. The dimensionless strength of the local nonlinearity is 
${\cal W}\approx 0.01\ll 1$, which is much smaller than the strength (in the order of 1) of the NNL, and will be neglected in the following. With this approximation and focusing on the diffraction along the $x$-axis, we convert the propagation equation in a dimensionless form, 
\begin{equation}\label{DE1}
i\frac{\partial u}{\partial \zeta}+\frac{1}{2}\frac{\partial^2 u}{\partial \xi^2}-\mathcal{V}u
+g_0\int d\xi'g(\xi',\xi)|u(\xi',\zeta)|^2u=0,
\end{equation}
where we have defined dimensionless quantities $u=\Omega_p/U_0$, $\xi=x/R_0$, $\zeta=z/L_{\rm diff}$, 
$\mathcal{V}=2L_{\rm diff}V_1$, $g=2L_{\rm diff}U_0^2R_0^2\int G({\bf r}_{\perp},{\bf r^{\prime}_{\perp}})dy'$, and $g_0=1/|\int g(\xi',\xi)d\xi'|$. The optical field and spatial coordinate have been scaled with respect to the maximal Rabi frequency $U_0$ and beam radius $R_0$.
The imaginary  potential $\mathcal{V}=iV_I$, where 
$V_I$ is gain (loss) when $V_I>0$ ($V_I<0$). 

Due to the Rydberg blockade, the response function has a soft-core shape [see Fig.~\ref{fig_model}(c)]. However the expression of  $g(\xi',\xi)$ is typically lengthy and complicated  (see Appendix~\ref{B} for  discussions). It can be approximated by an analytical form~\cite{hang2018PRA,Hang2023PRA}
\begin{align}\label{response function}
g(\xi',\xi)
\approx -\frac{B_1}{B_2\sigma^6+|\xi'-\xi|^6},
\end{align}
where $\sigma=R_b/R_0$ characterizes the nonlocal degree of the  nonlinearity. Here $R_b=|C_6/\delta_{\rm EIT}|^{1/6}$ is the radius of the blockade sphere with  $\delta_{\rm EIT}=|\Omega_c|^2/\Delta_3$ the linewidth of EIT transition spectrum.
$B_1$ and $B_2$ are the coefficients determined by laser parameters. 
The relation between $B_{1,2}$ and $b_{1,2}$ are $B_1=\sigma^6/b_2$ and $B_2=b_1/b_2$. 
As shown in Fig.~\ref{fig_model}(c), Eq.~(\ref{response function}) has a soft-core shape, which also agrees with the  numerical data. Note that this system has a \textit{defocusing}  nonlinearity as $g(\xi',\xi)<0$, which is crucial for the generation of shock waves.

 When $R_b$ is comparable to $R_0$, the nonlinear interaction is nonlocal (i.e. $\sigma$ is finite). In the opposite regime when $R_b\ll R_0$, we have a local regime as $\sigma\sim 0$.  The nonlocality parameter $\sigma$ can be varied by varying $R_b$ or $R_0$. The blockade radius can be tuned by changing detuning, laser intensities, or choosing different Rydberg states as $C_6\propto n^{11}$ with $n$ to be the principal quantum number. When $\sigma\sim 0$, we can make a local field approximation, i.e. $\int d\xi'g(\xi',\xi)|u(\xi',\zeta)|^2\approx |u(\xi,\zeta)|^2\int d\xi'g(\xi',\xi)$. Carrying out the spatial integration, we arrive at a propagation equation with local nonlinear interactions,
\begin{eqnarray}\label{eq:local NLSE}
i\frac{\partial u}{\partial \zeta}+\frac{1}{2}\frac{\partial^2 u}{\partial \xi^2}-\mathcal{V}u
+\bar{g}_0|u|^2u=0,
\end{eqnarray}
where $\bar{g}_0 =-g_0 \pi\sigma\left(b_1/b_2\right)^{1/6}/(3b_1)$ is the effective interaction strength. The local interaction is similar to the conventional Kerr nonlinearity, though the nonlinearity is much stronger. In the following, we will discuss the local and nonlocal regime separately.

\section{Local Rydberg nonlinearity regime}\label{sec4}
\subsection{Euler-like fluid equation}
In the local regime, we start  to investigate the generation of shock waves with the hydrodynamic approach. By treating the light field as a classical fluid, the hydrodynamic equation can be obtained by using the Madelung transformation $u(\xi,\zeta) = \sqrt{\rho(\xi,\zeta)}e^{i\phi(\xi,\zeta)}$, 
Eq.~(\ref{eq:local NLSE}) can be transformed into two Euler-like fluid equations,
\begin{subequations}\label{Euler}
\begin{align}
&\frac{\partial \rho}{\partial \zeta}+\frac{\partial}{\partial \xi}(\rho v)=2\rho V_I,\\
&\frac{\partial v}{\partial \zeta}+\frac{\partial}{\partial \xi}\left[\frac{1}{2}v^2+\bar{g}_0\rho+Q \right]
=0,\label{speed}
\end{align}
\end{subequations}
where $Q=-\frac{1}{2\sqrt{\rho}}\frac{\partial^2 \sqrt{\rho}}{\partial \xi^2}$ is quantum pressure,
and $v=\partial \phi/\partial \xi$ is the flow velocity of the light fluid. 
Neglecting $Q$ for the moment, Eq.~(\ref{speed}) becomes
\begin{align}
\label{eq:no_qp}
\frac{\partial v}{\partial \zeta}+\frac{\partial }{\partial \xi}\left(\frac{1}{2}v^2+\bar{g}_0\rho \right)=0.
\end{align}
Equations~(\ref{Euler}a) and (\ref{eq:no_qp}) can be cast into the diagonal Riemann form
\begin{subequations}\label{diagonal Riemann equation}
\begin{align}
&\frac{\partial r_1}{\partial \zeta}+c_1\frac{\partial r_1}{\partial \xi}=d_1,\\
&\frac{\partial r_2}{\partial \zeta}+c_2\frac{\partial r_2}{\partial \xi}=d_2,
\end{align}
\end{subequations}
where the Riemann invariant and hyperbolic speeds are defined by 
$r_i={v}/{2}\pm \sqrt{{\bar g}_0\rho}$ and $c_i=v\pm \sqrt{\bar{g}_0\rho}$,
with $c_1=(3r_1+r_2)/2$, and $c_2=(r_1+3r_2)/2$. $d_{1,2}=\pm (r_1-r_2)V_I/2$.
The light fluid intensity and the flow velocity are given by
$\rho=(r_1-r_2)^2/(4\bar{g}_0)$ and $v=r_1+r_2$.

\begin{figure}
\centering
\includegraphics[width=1\linewidth]{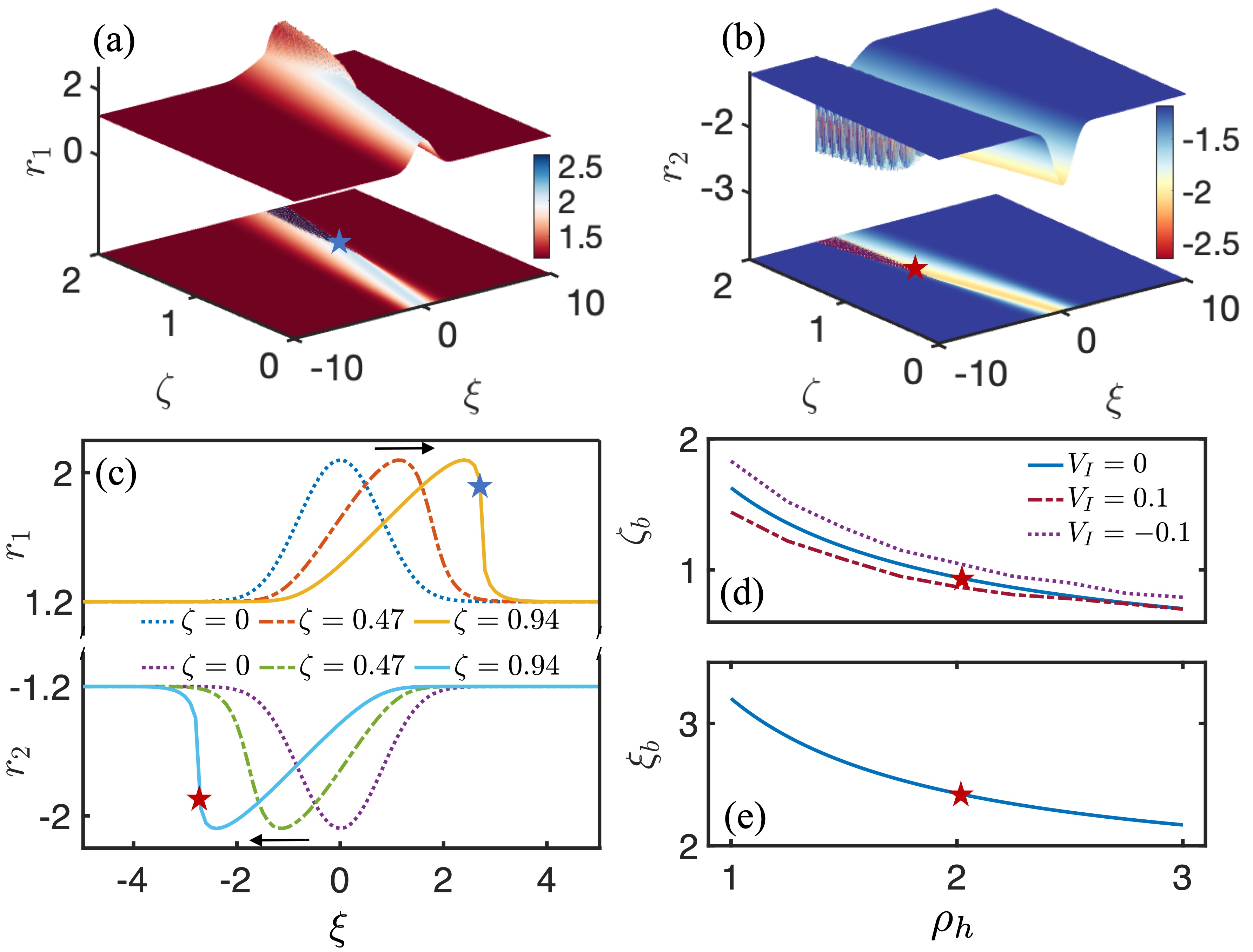}
\caption{(a) Right-moving and (b) left-moving Riemann waves. The stars represent the  breaking position of the wave, after which the shock wave forms.
(c) Riemann waves $r_1$ and $r_2$ as function as $\xi$ for $\zeta=0,\,0.47,\,0.94$. The stars represent the  breaking position.  Upper panel: The right-moving part $r_1$. Lower panel: The left-moving part $r_2$. 
(d) Breaking point in the $\zeta$ direction ($\zeta_b$) as function as the hump peak intensity $\rho_h$. The star marks the breaking point corresponding to panel (a) and (b).
(e)~The same as (d), but for breaking point in the $\xi$ direction ($\xi_b$). Other parameters are $\rho_b=1$, $\rho_h=2$, $\xi_0=1$, $\bar{g}_0=-1.44$, and $V_I=0$. In panel (d), we additionally show data with dissipative potentials. 
}
\label{fig_Riemann_wave}
\end{figure}

We numerically solve the Riemann equation without dissipative potential with the initial condition
\begin{align}\label{Rho0}
\rho(\xi,0)=\rho_b+\rho_h e^{-\xi^2/\xi_0^2},~\text{and} ~v(\xi,0)=0.
\end{align}
Here $\rho_b$ and $\rho_h$ are the background and hump intensity, and $\xi_0$ is the width of the hump. Figure~\ref{fig_Riemann_wave}(a) and (b) are the propagation of the right and left-moving Riemann waves, respectively.
The wave is stable before the shock onsets. After a critical distance (marked by stars on the figure) the shock forms. A distinctive feature is that the wave becomes steepening at the critical distance. These points are the shock wave breaking points  (see more discussions in Sec.~\ref{sec:breakingpoint}). These points are symmetric for the left and right-moving solutions, as depicted in Fig.~\ref{fig_Riemann_wave}(c). In other words the left and right-moving components will form shock waves after propagating equal distances. 

\begin{figure*}[ht]
\centering
\includegraphics[width=0.93\linewidth]{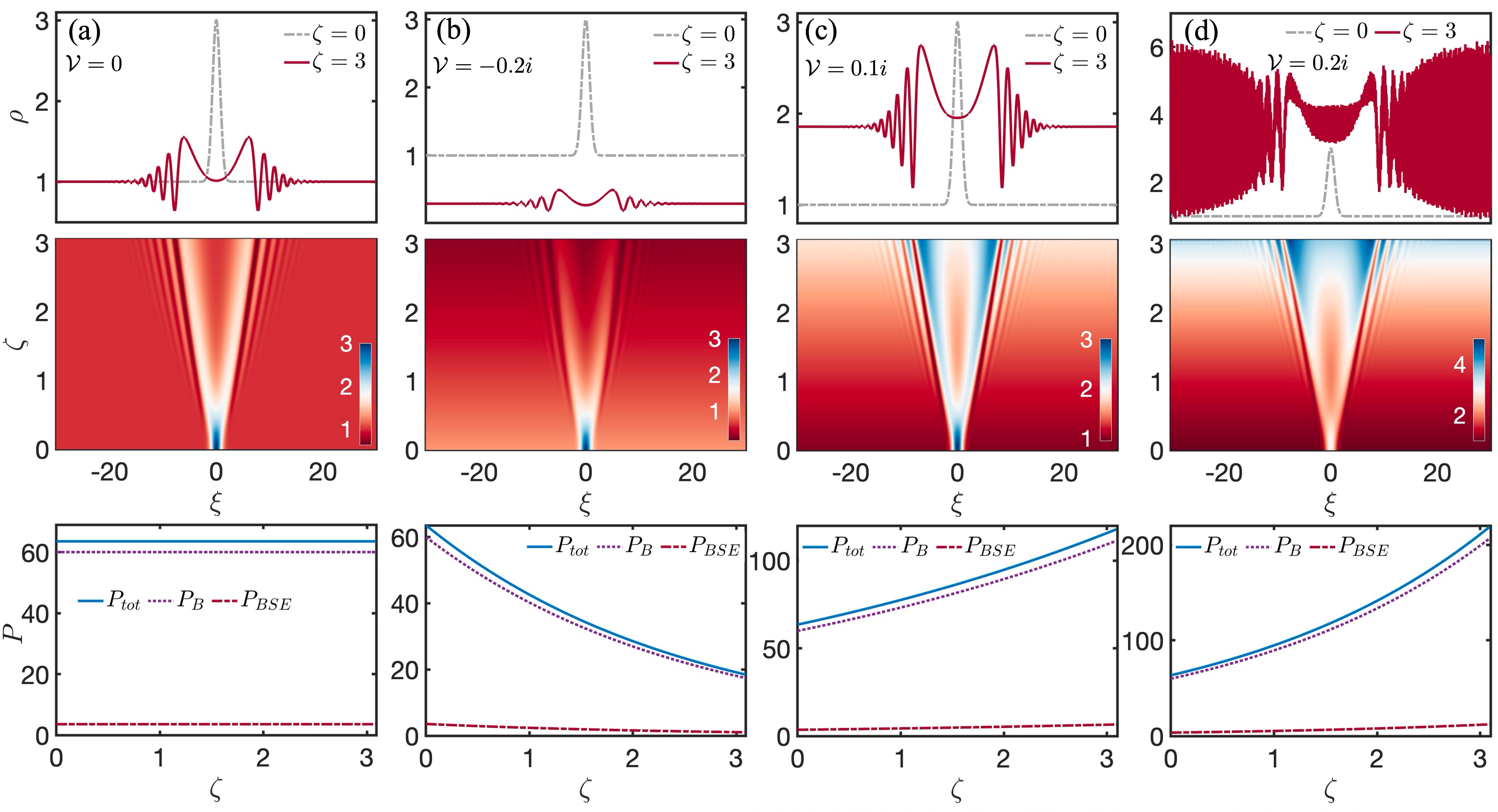}
\caption{
(a), (b), (c), and (d) show the propagation of shock waves with $\mathcal{V}=0$ , $\mathcal{V}=-0.2i$ (loss), $\mathcal{V}=0.1i$, and $\mathcal{V}=0.2i$ (gain). ~Upper panel: The probe field intensity $\rho=|u|^2$ at $\zeta=0$ and $3$ (dotted-dashed grey and solid red lines). Middle panel: Wave propagation from $\zeta=0$ to $\zeta=3$. Lower Panel: The total power of the probe field ($P_{tot}$, blue solid line), the background ($P_B$, dotted purple line) and exchange ($P_{BSE}$, dotted-dashed red line) power.  In all panels, the initial conditions are $\rho_b=1$, $\rho_h=2$, and $\xi_0=1$. 
}
\label{fig_propagation_local}
\end{figure*}

\subsection{Breaking point}
\label{sec:breakingpoint}
In the local regime and without dissipation, the breaking point can be found analytically. We first linearize  Eq.~(\ref{diagonal Riemann equation})  by means of the hodograph transform (see, Refs.~\cite{kamchatnov2000book,ivanov2020PRE,isoard2019PRA}), which treats $\xi$ and $\zeta$ as functions of $r_1$ and $r_2$. The transformation yields,
\begin{align}\label{hodograph transform}
\frac{\partial \xi}{\partial r_1}-c_2\frac{\partial \zeta}{\partial r_1}=0,~~~\frac{\partial \xi}{\partial r_2}-c_1\frac{\partial \zeta}{\partial r_2}=0.
\end{align}
We then introduce two functions $w_1(r_1,r_2)$ and $w_2(r_1,r_2)$ such that
\begin{align}\label{xi_c}
\xi-c_1\zeta=w_1,~~~~\xi-c_2 \zeta=w_2.
\end{align}
Using the initial condition, $w_1$ and $w_2$ can be obtained,
\begin{align}
w_{1,2}=
\begin{cases}
\xi_0\sqrt{{\rm ln}{\rho_h}-{\rm ln}[r_{1,2}^2/\bar{g}_0-\rho_b]},~&\xi>0,\\
-\xi_0\sqrt{{\rm ln}{\rho_h}-{\rm ln}[r_{1,2}^2/\bar{g}_0-\rho_b]},~& \xi<0.
\end{cases}
\end{align}
Wave breaking corresponds to the occurrence of a gradient catastrophe for which $|\partial r_{1,2}/\partial \xi|\rightarrow \infty$. As the right and left-moving wave are symmetric with respect to the $\xi$, we determine the breaking point of the right-moving branch explicitly. From Eq.~(\ref{xi_c}), wave breaking occurs at a distance $\zeta$ such that~\cite{isoard2019PRA}
\begin{align}\label{zbreaking1}
\zeta=\frac{2}{3}
\left|\frac{\partial w_1}{\partial r_1}\right|_{r_1=r_1^*}.
\end{align}
One can evaluate $\zeta_b$ approximately when the point of largest gradient in $r_1(\xi)$ lies in a region where $r_2 \approx \sqrt{\bar{g}_0\rho_b}$. In this case, Eq.~(\ref{zbreaking1}) becomes
\begin{align}\label{zbreaking2}
\zeta_b \approx  \frac{ 2\xi_0\sqrt{\rho^*}}{3\sqrt{\bar{ g}_0}(\rho^*-\rho_b)\sqrt{\ln[\rho_h/(\rho^*-\rho_b)]}}.
\end{align}
The shortest of distance $\zeta$ is reached close to the point $\xi^*$ for which $|\partial \rho/\partial \xi|$ is maximal. We denote $\xi^*$ the coordinate of this point in the $\xi$ direction and $\rho^*=\rho(\xi^*)$. According to $\partial \xi/\partial \rho=\partial^2 \xi/\partial \rho^2=0$ at ${\rho=\rho^*}$, it is readily to find
$\ln\left[\rho_h/(\rho^*-\rho_b )\right]=\rho^*/(\rho^*+\rho_b)$. This leads to the approximate relation,
\begin{align}\label{wave-breaking distance}
\zeta_b \approx \frac{2\xi_0}{3(\rho^*-\rho_b)}\sqrt{\frac{\rho^*+\rho_b}{\bar{g}_0}}.
\end{align}
The breaking point as a function of $\rho_h$  is shown in the Fig.~\ref{fig_Riemann_wave}(d), which matches the numerical calculation well. The breaking point $\zeta_b$ is reduced when increasing the hump intensity. Such relation is useful in controlling the generation of shock waves. For example, increasing the intensity of the hump peak allows for a shorter distance and faster visibility of the shock wave.  Moreover, the breaking point $\xi_b$ along the $\xi$ axis when the wave breaks along the $\zeta$ direction can be obtained~\cite{isoard2020EPL},
\begin{align}\label{xbreaking}
	\xi_b\approx c_s(\rho^*)\zeta_b+\xi_0\sqrt{{\rm ln}{\rho_h}-{\rm ln}[\rho^*-\rho_b]}.
\end{align}
Here $c_s=\sqrt{\bar{g}_0\rho}$ is the local sound speed. The results are shown in Fig.~\ref{fig_Riemann_wave}(e), which agrees with the numerical calculation well. 

Including dissipation in the Riemann function, analytical solutions are in general not possible. Instead, we find the breaking point numerically. Breaking points for  $V_I=0.1$ and $-0.1$ are shown in Fig.~\ref{fig_Riemann_wave}(d).  It is found that the breaking point $\zeta_b$ decrease as the dissipative potential changes from loss to gain. These results highlight the importance of dissipative potential on the generation of shock waves. For example, the gain potential (i.e. $V_I>0$) accelerates the generation of shock waves, as the wave breaks earlier. The loss potential ($V_I<0$) slows down their generation.

On the other hand, the nonlocality can also affect the breaking point. To be specific, the breaking point $\zeta_b$ increases with $\sigma$. Therefore, a strong nonlocality will postpone the occurrence of shock waves~\cite{ghofraniha2007PRL}.


\subsection{Wave propagation}
We now turn to investigate the propagation of shock waves by numerically solving Eq.~(\ref{eq:local NLSE}), where the quantum pressure is taken into account explicitly. In Fig.~\ref{fig_propagation_local} propagation of shock waves without external potential [Fig.~\ref{fig_propagation_local}(a)], in loss potential [Fig.~\ref{fig_propagation_local}(b)], and in gain potential [Fig.~\ref{fig_propagation_local}(c)-(d)] are shown. Without dissipation, the initial hump  splits into two density peaks firstly. When the shock wave forms, the wave front  oscillates rapidly in the $\xi$ direction, as depicted in the upper and middle panel of Fig.~\ref{fig_propagation_local}(a). The end of oscillations edge corresponds to small-amplitude edge of the shock wave~\cite{kamchatnov2019PRE,Hang2023PRA}. Before the shock wave reaches the boundary, the background field is not perturbed.  

In a loss potential ($\mathcal{V}=-0.2i$), the intensity of the background wave decays exponentially, as shown in Fig.~\ref{fig_propagation_local}(b). Shock waves form in the central region, characterized by the rapid oscillation at the shock edge. Amplitudes of the oscillation edge 
become smaller than that of the $\mathcal{V}=0$ case.  Results of shock waves in a gain potential with $V_I=0.1$ are shown in Fig.~\ref{fig_propagation_local}(c). From the upper panel of Fig.~\ref{fig_propagation_local}(c), the intensity of the shock wave and background field, and the small-amplitude edge are all lager than the cases $V_I=0$ and $V_I=-0.2$. This is a direct manifestation of the gain effect. On the other hand, further increasing the strength of the gain potential, the shock wave and background field quickly become unstable, causing catastrophic collapse~\cite{Hang2023PRA}.  

In the presence of the dissipative potential, power of the field will decay or grow exponentially with the propagation distance $\zeta$. The total, background, and exchange power between shock wave and background fields are obtained,
$ P_{tot}=\int |u(\xi,\zeta)|^2 d\xi\approx e^{2V_I\zeta}\int|u(\xi,\zeta=0)|^2d\xi$, 
$P_B=\int |u_b(\xi,\zeta)|^2d\xi\approx e^{2V_I\zeta}\int \rho_b d\xi$,
and $P_{BSE}=P_{tot}-P_B\approx e^{2V_I\zeta}\int\rho_h\exp(-\xi^2/\xi_0^2)d\xi$. Here $u_b(\xi,\zeta)=\sqrt{\rho_b}\exp(V_I\zeta)$ is the boundary intensity at $(\xi_{L},\zeta)$. When $V_I=0$, the total, background as well as the exchange power is conserved as a function of $\zeta$ [lower panel of Fig.~\ref{fig_propagation_local}(a)]. Their values are determined by the initial values. On the other hand, the power will grow (decay) exponentially when $V_I>0$ ($V_I<0$) when propagating in the medium, as shown in the lower panel of Fig.~\ref{fig_propagation_local}(b) and (c). 

\begin{figure}[t]
\centering
\includegraphics[width=1\linewidth]{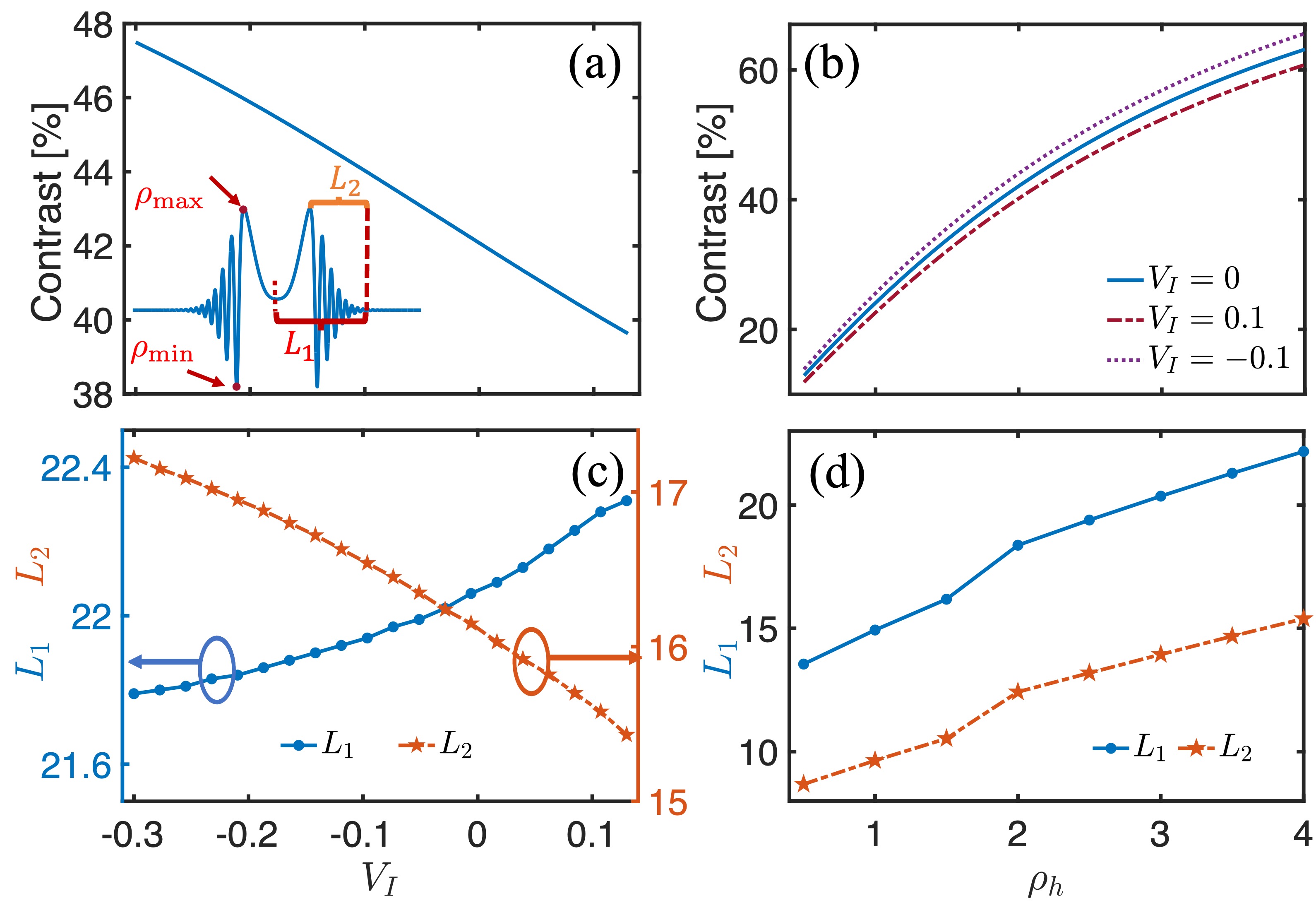}
\caption{
(a) Oscillation contrast versus $V_I$.  The insert illustrate the maximum and minimum values of $\rho$, i.e., $\rho_{\rm max}$ and $\rho_{\rm min}$, shock width $L_1$, and oscillation width $L_2$. The parameters $\rho_b=1$, $\rho_h=2$, $\xi_0=1$, $\bar{g}_0=-1.44$, and $\zeta=3$.  
(b) Contrast versus $\rho_h$ with $V_I=0,\,0.1$, and $-0.1$, respectively.
(c) Shock width $L_1$, measured from the center to the end of oscillations, with respect to the intensity of imaginary potential $V_I$ (blue solid line). 
The oscillation width $L_2$, measured from the start to the end of oscillations (orange dotted-dashed line). 
(d) $L_1$ and $L_2$ by varying the hump peak intensity $\rho_h$ with $V_I=0$.
The other parameters same as panel~(a).}
\label{fig_contrast_local}
\end{figure}

\subsection{Contrast and shock width} 
\label{sec:contrast}
As shown in Fig.~\ref{fig_propagation_local}, profiles of the shock wave exhibit a non-trivial dependence on the dissipative potential and initial state. Once the shock wave forms, rapid oscillations are found along the $\xi$ axis. Including the quantum pressure, the gradient divergence of $u$ is not available, which makes it impossible to calculate the breaking point. As shock waves oscillate rapidly, the maximal and minimal values of the oscillations provide a way to characterize the amplitude of the shock wave. Therefore we calculate the visibility of the oscillations near the soliton edge (i.e., the start of oscillation edge) of the shock wave by measuring the contrast~\cite{xu2016OL,isoard2019PRA}
\begin{align}
\mathcal{C}=\frac{\rho_{\rm max}-\rho_{\rm min}}{\rho_{\rm max}+\rho_{\rm min}},
\end{align}
where $\rho_{\rm max}$ and $\rho_{\rm min}$ are the maximum and minimum values of $\rho$, as depicted in the left lower insert of Fig.~\ref{fig_contrast_local}(a). At a fixed propagation distance $\zeta$, $\mathcal{C}$ as a function $V_I$ is shown in Fig.~\ref{fig_contrast_local}(a).
As $V_I$ increases (from loss to gain), the contrast of the formed shock wave decreases.
In Fig.~\ref{fig_contrast_local}(b), we show contrast $\mathcal{C}$ when varying $\rho_h$.
For a given $V_I$, the contrast of the shock wave increases with increasing $\rho_h$. Changing $V_I$, such trend remains the same. These results show that we could enhance the contrast by using larger $\rho_h$.

We also calculate the shock wave width $L_1$ (measured from the center to the end of the intensity oscillation)~\cite{wan2007NP,barsi2007OL}, and the oscillation width $L_2$ (measured from the start to the end of the oscillation), as indicated in the insert of Fig.~\ref{fig_contrast_local}(a). We find $L_1$ increases with $V_I$, as shown in Fig.~\ref{fig_contrast_local}(c). The reason is that the small-amplitude edge has a slight increase when the potential changes from loss to gain.
However, the width of oscillation $L_2$ (orange solid line) decreases due to the soliton edge increases. 

In Fig.~\ref{fig_contrast_local}(d) width $L_1$ and $L_2$ as function of the hump peak intensity $\rho_h$ are shown. 
The larger the hump density $\rho_h$, the wider the width $L_1$ and $L_2$. We can understand these distances by examining the local sound speed $c_s\propto \sqrt{\rho}$. $c_s$ is a function of local density $\rho$ that consists of both the background and hump density.  Increasing the hump density will increase the local sound speed. This means $L_1$ will be larger with higher $\rho_h$, after propagating certain $\zeta$. The inner region defined by $L_2$, on the other hand, describes propagation of solitons~\cite{Hang2023PRA}. Its front travels at the sound speed approximately. Hence $L_2$ increases when $c_s$ ($\rho_h$) is larger.

\section{Nonlocal Rydberg nonlinearity regime}\label{sec5}

We will consider the full soft-core potential using Eq.~(\ref{response function}). To understand the role played by the nonlocality, we will solve Eq.~(\ref{DE1}) numerically by taking into the full soft-core potential Eq.~(\ref{response function}). There are two different ways to change the soft-core potential Eq.~(\ref{response function}). One can vary $\sigma$ by fixing $B_1$ and $B_2$, which requires to change parameters $b_1$ and $b_2$ correspondingly. The depth of the potential [see Fig.~\ref{fig_model}(c)]
is $B_1/(B_2\sigma^6) = 1/b_1$. As $b_1=B_2\sigma^6/B_1$, the depth of the potential increases rapidly as $\sigma$ decreases, which eventually makes the numerical calculation unstable. Hence $\sigma$ can not be too small in practice. To avoid this divergence, one can alternatively change $\sigma$ while keeping the potential depth (hence $b_1$) constant. 
\begin{figure}[t]
\centering
\includegraphics[width=0.93\linewidth]{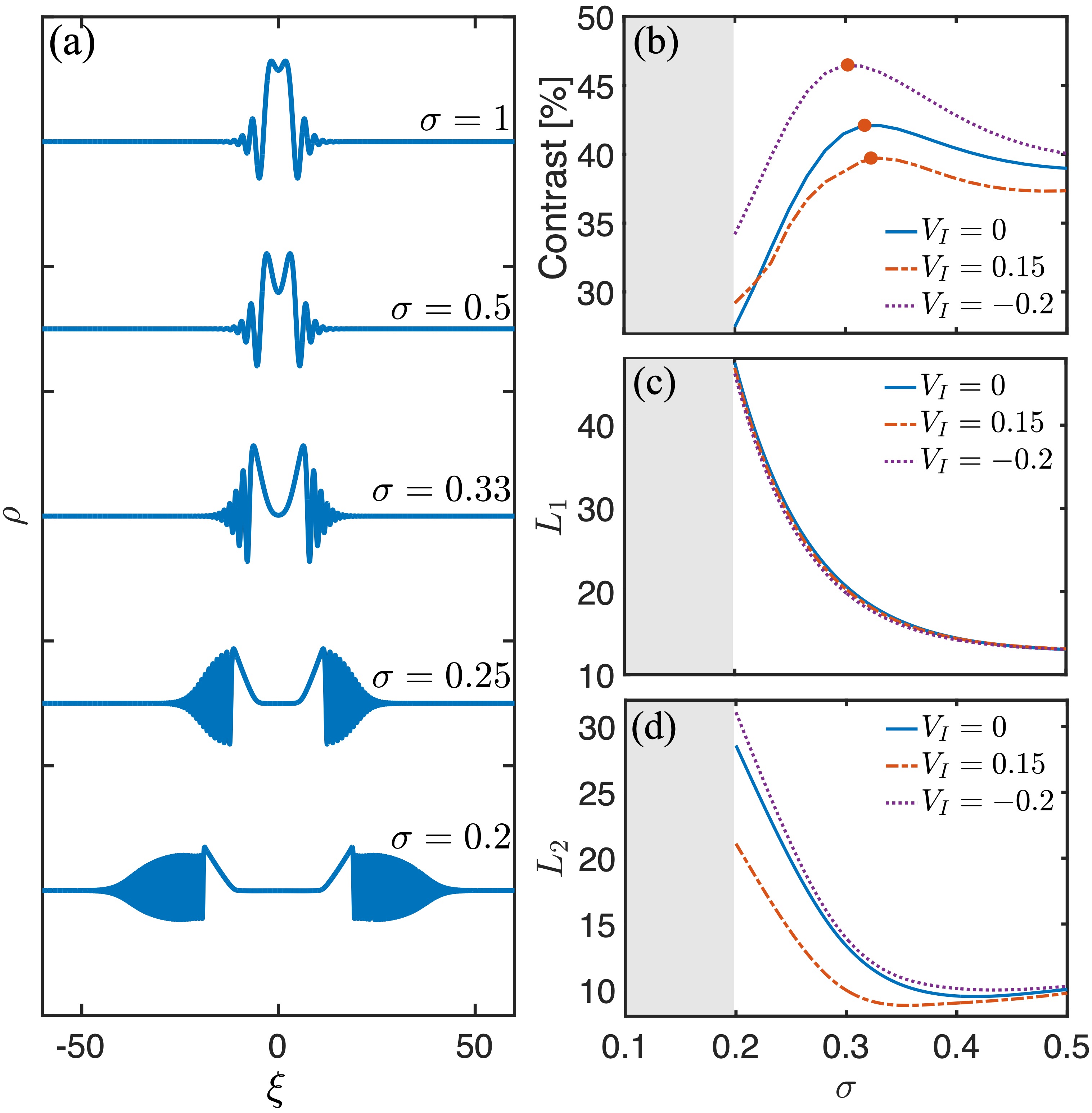}
\caption{(a) Shock waves for different nonlocality degree  $\sigma=0.2,\,0.25,\,033,\,0.5,\,\text{and}\, 1$ (from bottom to top) with $V_I=0$ and $\zeta=3$. 
(b) The oscillation contrast as function of the nonlocality degree $\sigma$ with $V_I=0,\,0.15,\,\text{and}\,-0.2$ at $\zeta=3$. The red dot represents the maximal contrast.
(c) The width $L_1$ and (d) oscillation width $L_2$ as function of the nonlocality degree $\sigma$ with $V_I=0,\,0.15,\,\text{and}\,-0.2$ at $\zeta=3$. The initial condition $\rho_h=2$, $\rho_b=1$, and $\xi_0=1$. 
}
\label{fig:propagation1}
\end{figure}

\subsection{Fix $B_1$ and $B_2$}
\label{sec:nonlocalA}
To avoid numerical instabilities, we have chosen the smallest $\sigma=0.2$ in the numerical simulation. When $\sigma$ is small, the formation of shock waves are featured by conspicuous oscillations, as illustrated in Fig.~\ref{fig:propagation1}(a). The behavior is similar to that of the local nonlinearity regime. Increasing $\sigma$, the oscillation 
frequency decreases. As a result,  
the contrast increases rapidly with increasing $\sigma$, and reaches its maximum value $\mathcal{C}\approx0.47$ around $\sigma\approx 0.33$, as shown in Fig~\ref{fig:propagation1}(b). It subsequently decreases and saturates to a constant value. The width $L_1$ and $L_2$ depend on $\sigma$ sensitively when $\sigma$ is small. They decrease rapidly with increasing $\sigma$, as shown in Fig.~\ref{fig:propagation1}(c) and (d). This dependence can be understood by analyzing the sound speed. 
When $B_1$ and $B_2$ are fixed, we obtain $c_s=\sqrt{\pi g_0 \rho B_1/(3B_2^{5/6}\sigma^5})$. The sound speed decreases rapidly ($c_s\propto 1/\sqrt{\sigma^5}$) when $\sigma$ increases. Therefore both $L_1$ and $L_2$ reduce when $\sigma$ is large.

When increasing $\sigma$, we in fact drive the response from highly nonlinear to a linear regime. In other words, when $\sigma$ is small, the wave dynamics is strongly nonlinear, which promotes the generation of shock waves. By increasing $\sigma$, however, the response of the Rydberg medium becomes effectively linear. When $\sigma$ ($R_b$) is large, the soft-core potential is nearly a constant compared to the typical wavelength of the excitation. Assuming the nonlocal potential is a constant, one can carry out the integration in Eq.~(\ref{DE1}) and obtain a linear potential~\cite{barsi2007OL}, i.e. $\int d\xi'g(\xi',\xi)|u(\xi')|^2u(\xi)\approx g(0,0) P_{tot} u(\xi)$. As a result, the resulting wave will propagate linearly, i.e. behaves like phonons (see  Fig.~\ref{fig_propagation_withcs} in Appendix).
We will focus on shock wave generation and propagation in the nonlocal regime. In practice, this requires roughly $\sigma<0.5$, i.e. the blockade radius is half of $R_0$. When $\sigma>0.5$, the generated wave is linear and show similar propagation dynamics.
\begin{figure}[t]
\centering
\includegraphics[width=1\linewidth]{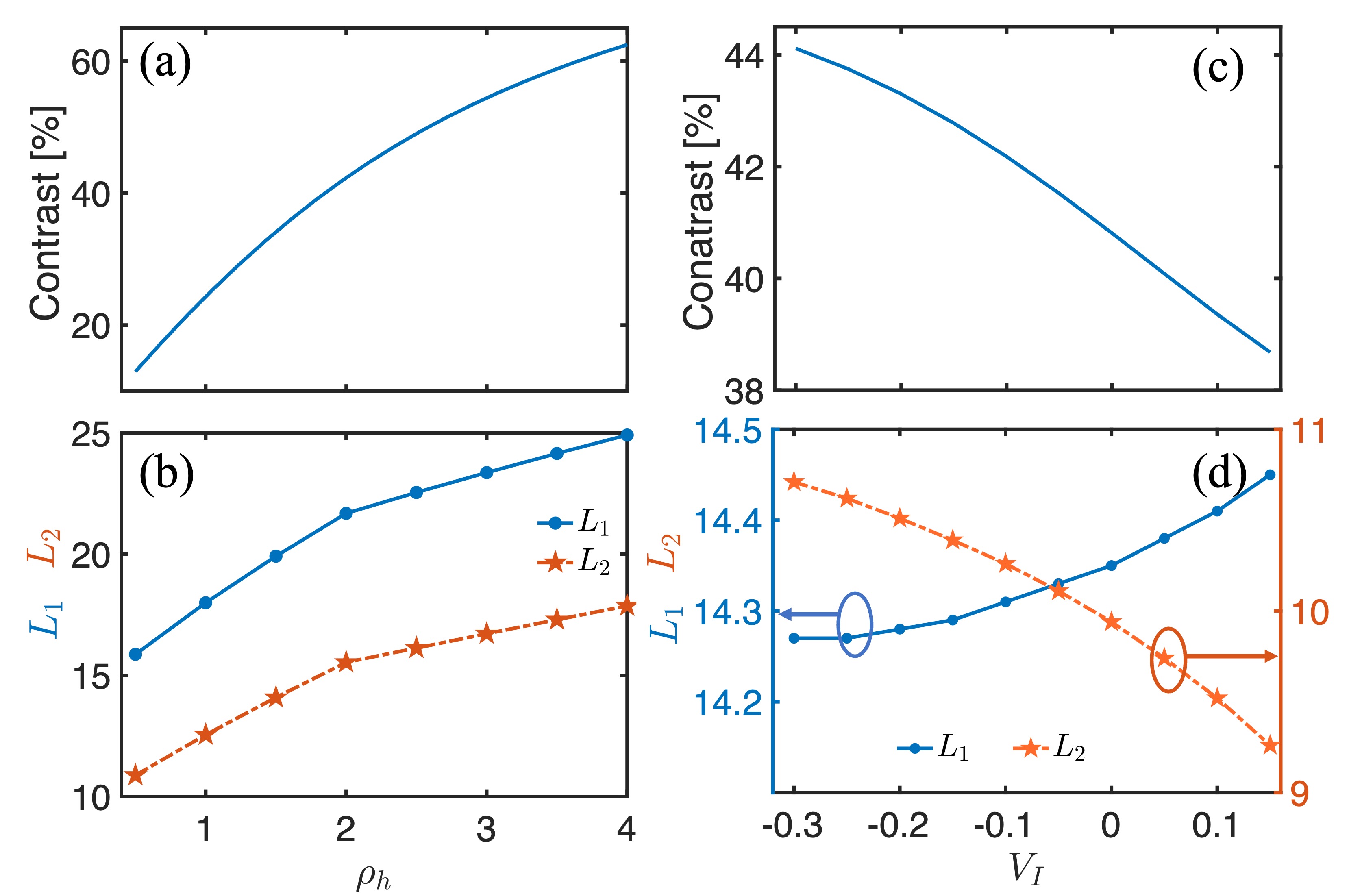}
\caption{(a) Oscillation contrast and (b) width $L_1$ and $L_2$ versus $\rho_h$
  with $\sigma=0.33$ and $V_I=0$.  
(c) Oscillation contrast and (d) width $L_1$ and $L_2$ versus $V_I$ with $\sigma=0.39$, $\rho_h=2$ at $\zeta=3$. The other parameters same as Fig.~\ref{fig_contrast_local}(a).
}
\label{fig:contrast nonlocal1}
\end{figure}

When increasing $\rho_h$, the contrast increases monotonically, as   shown in Fig.~\ref{fig:contrast nonlocal1}(a).  Both $L_1$ and $L_2$ become larger for higher $\rho_h$, as $L_1,\,L_2\propto c_s\propto \sqrt{\rho_h}$. Similar dependence are also found in the local nonlinear case [see Fig.~\ref{fig_contrast_local}(b) and (d)]. The difference is that both $L_1$ and $L_2$ are slightly larger in the nonlocal Rydberg medium than that of the local medium (for given $\rho_h$), mainly due to that the strength of the nonlocal interaction is different in the two figures.

We then examine the characteristic quantities as a function of $V_I$ numerically. The results are shown in Fig.~\ref{fig:contrast nonlocal1}(c) and (d). When varying $V_I$, the contrast, $L_1$ and $L_2$ exhibit similar dependence on $V_I$ as found in the local nonlinearity case. Changing $V_I$ from $-0.3$ to $0.15$, the contrast decreases slowly, as shown in Fig.~\ref{fig:contrast nonlocal1}(c), akin to the finding in the local regime, as shown in Fig.~\ref{fig_contrast_local}(a). Compared to the local case, shock width $L_1$ is barely change when increasing $V_I$,  as shown in Fig.~\ref{fig:contrast nonlocal1}(d). The weak dependence comes from the fact that the small-amplitude edge only has a slight change. Oscillation width $L_2$ decreases apparently, due to the soliton edge increases, similar to the local case shown in Fig.~\ref{fig_contrast_local}(c). The numerical data show that the contrast, $L_1$ and $L_2$ are all smaller than that of the local nonlinear case.

\subsection{Fix the potential depth}\label{sec:nonlocalB}

\begin{figure}[t]
\centering
\includegraphics[width=0.93\linewidth]{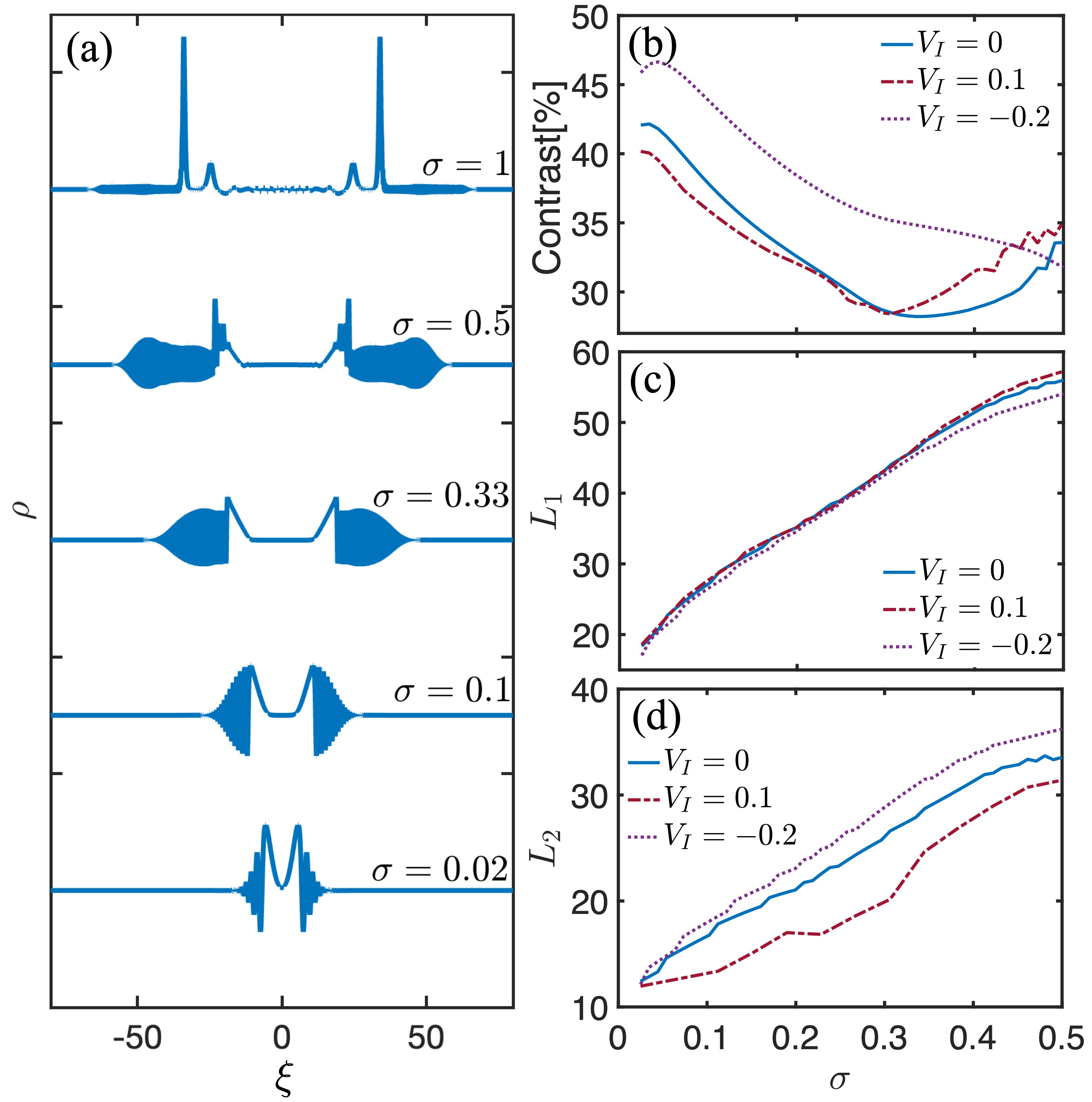}
\caption{(a) Shock waves for different nonlocality degree $\sigma=0.02,\,0.1,\,0.33,\,0.5$, and $1$, with $b_1=0.022$, $b_2=0.5$, $V_I=0$, and $\zeta=3$. (b) The oscillation contrast of the shock wave as function of the nonlocality degree $\sigma$ with $V_I=0,\,0.1,\,\text{and}\,-0.2$ at $\zeta=3$. 
(c) The shock width $L_1$ and (d) oscillation width $L_2$ as a function of $\sigma$ with $V_I=0,\,0.1,\,\text{and}\, -0.2$ at $\zeta=3$. The initial condition $\rho_h=2$, $\rho_b=1$, and $\xi_0=1$. 
}
\label{fig:propagation2}
\end{figure}

In this case, $b_1$ is a constant. The sound speed $c_s\propto \sqrt{\sigma\rho}$, which means the larger $\sigma$, the more separation between the left and right moving shock wave, as shown in Fig.~\ref{fig:propagation2}(a). When $\sigma=0.5$ and 1, we find high density peaks at in the inner region, which result from that the system is effectively linear. They also affect the contrast, shown in Fig.~\ref{fig:propagation2}(b). It decreases gradually, arrives at a minimum, and then increases again with increasing $\sigma$. For large $\sigma$, the rising contrast is purely caused by the inner peaks and the background density, where the amplitude of the shock wave is marginal. Moreover, both $L_1$ and $L_2$ increase monotonically as we increase $\sigma$, due to $c_s\propto \sqrt{\sigma\rho}$. Thus after propagating distance $\zeta$, the left- and right-moving shock waves separately significantly.

Dissipation, on the other hand, leads to a nearly global shifts to the contrast. As shown in Fig.~\ref{fig:propagation2}(b), the contrast becomes larger when $V_I$ is negative (loss potential). When $V_I$ is positive, the contrast is only shifted lower slightly when $\sigma<0.3$. The dissipation merely modifies $L_1$ as we increase $\sigma$, which shows the robustness of the shock wave propagation. On the contrary, non-negligible changes to $L_2$ are found when $V_I$ is nonzero. This means that the speed of the soliton varies apparently. Compared to results shown in Fig.~\ref{fig:propagation1}(b)-(d), one notes that modifications of these quantities due to $V_I$ are identical in the two cases.

\begin{figure}[t]
\centering
\includegraphics[width=1\linewidth]{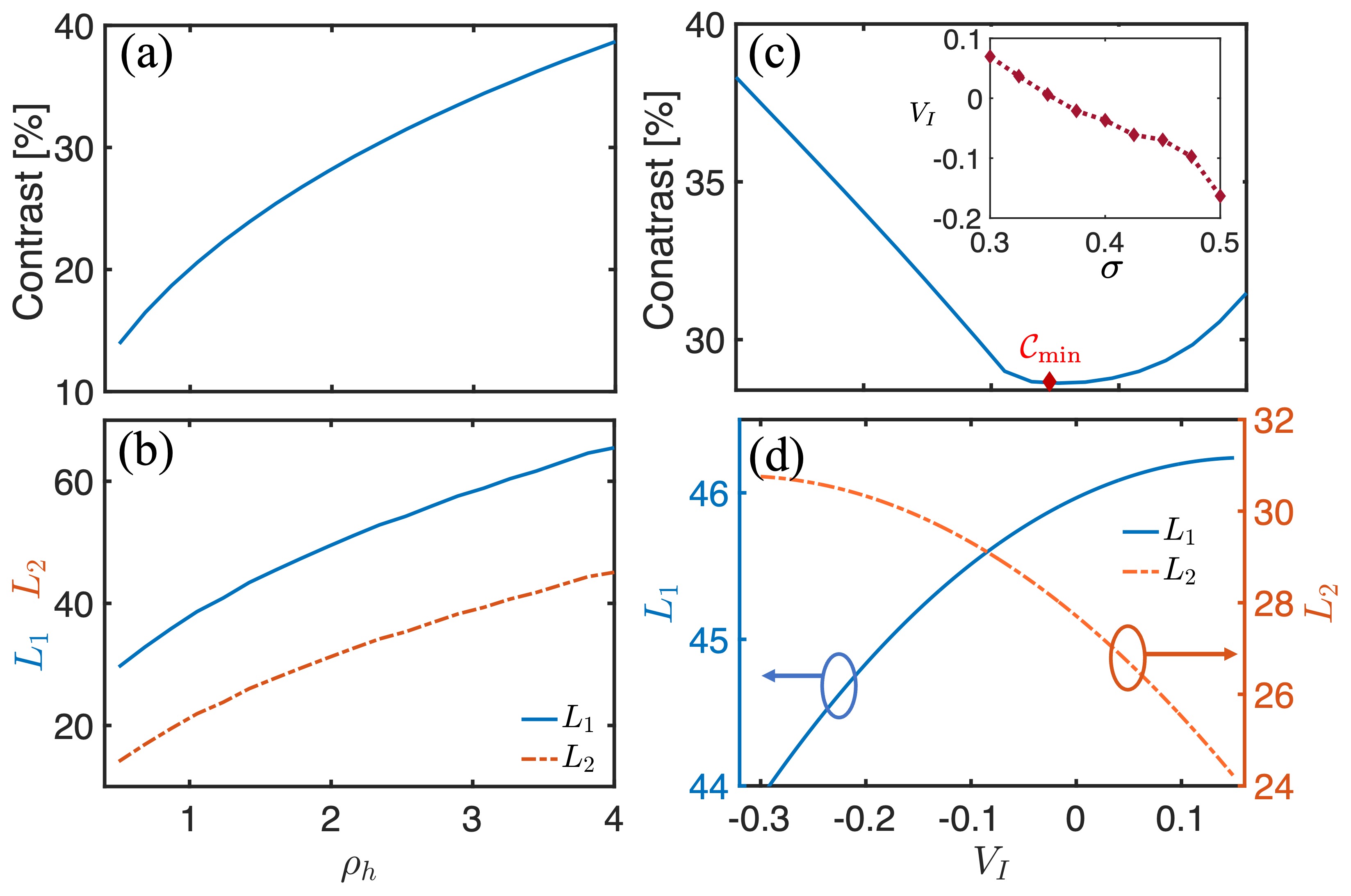}
\caption{
(a) Oscillation contrast and (b) Shock width $L_1$  and oscillation width $L_2$   versus  $\rho_h$
  with $\sigma=0.33$ and $V_I=0$ at $\zeta=3$.  
(c) Oscillation contrast and (d) Shock width $L_1$  and oscillation width $L_2$ versus $V_I$ with $\sigma=0.33$, $\rho_h=2$ at $\zeta=3$. The other parameters same as Fig.~\ref{fig:propagation1}. In panel (c), the inset illustrate the minimal contrast value $\mathcal{C}_{\rm min}$ versus $\sigma$ and $V_I$.
}
\label{fig:contrast nonlocal2}
\end{figure}

When increasing $\rho_h$, the contrast, $L_1$ and $L_2$ all increase [see Fig.~\ref{fig:contrast nonlocal2}(a) and (b)]. This is because not only the sound speed, but also amplitudes of the oscillation increases with larger $\rho_h$. Similar trends are also found in the previous case shown in Fig.~\ref{fig:contrast nonlocal1}(a) and (b). However, the contrast depends on $V_I$ non-trivially in the current case. By increasing $V_I$, we find contrast has a minimal value $\mathcal{C}_{\text{min}}$ [see Fig.~\ref{fig:contrast nonlocal2}(c)], which is different from the situation shown in Fig.~\ref{fig:contrast nonlocal1}(c). In the latter case, the contrast declines monotonically with increasing $V_I$ in the given parameter range. $\mathcal{C}_{\text{min}}$ depends on not only $V_I$, but also $\sigma$. We numerically obtain $\mathcal{C}_{\text{min}}$ and the respective parameter $V_I$ and $\sigma$. In the inset of Fig.~\ref{fig:contrast nonlocal2}(c), the corresponding $V_I$ and $\sigma$ are plotted. It shows that when $\sigma$ increases, one has to decreases $V_I$ in order to find $\mathcal{C}_{\text{min}}$.  Finally, $L_1$ and $L_2$ as a function of $V_I$ are shown in Fig.~\ref{fig:contrast nonlocal2}(d). The trend is similar to that of the previous case [Fig.~\ref{fig:contrast nonlocal1}(d)]. Their values are, however, larger in general. This results from the fact that the sound speed has different dependence on parameters in the two cases. 

\section{Conclusion}\label{sec6}
In this paper, we have elaborated a scheme that enables the generation and propagation of shock waves within an atomic gas involving a homogeneous dissipative potential and long-range Rydberg interaction under the condition of EIT. 
We have demonstrated that the homogeneous gain or loss potential significantly alters the power in the local nonlinearity regime. Both the oscillation contrast of shock waves and the oscillation width exhibit monotonic decreases as the potential changes from loss to gain. 
Furthermore, we have shown that in the NNL regime, the nonlocal degree of the nonlinearity modifies the oscillation amplitude and width of shock waves. Additionally, we have illustrated the hump intensity of the initial state can enhance the visibility of shock waves in both local and NNL regimes. Our results reveal the nontrivial roles of dissipation and nonlocality in the generation of shock waves, providing new routes to manipulate their profiles and stability. Our study opens new avenues for exploring non-Hermitian dynamics~\cite{moiseyev2011,bender2007,ashida2020,okuma2023}, and nonlinear wave dynamics~\cite{bai_self-induced_2020,RotschildPRL2005,rotschild2006NP} modulated by the interplay between the NNL, and local and nonlocal~\cite{yan_electromagnetically_2020,wuFastSpinSqueezing2022} dissipation in highly controllable Rydberg gases.

\acknowledgments 
We thank Zeyun Shi for helpful discussions.
L.Q., C.H, and G.H acknowledge the National Natural Science Foundation of China (NSFC) under Grants No.~12247146, No.~12374303, and No.~11975098, Henan Province Postdoctoral Research Funding Project, Overseas Training Program for High-level Innovative Talents in Henan Province's `Double First Class' Universities, and Shanghai Municipal Science and Technology Major Project under Grant No.~2019SHZDZX01. 
W.L. acknowledges support from the EPSRC through Grant No.~EP/W015641/1, the Going Global Partnerships Programme of the British Council (Contract No.~IND/CONT/G/22-23/26), and the International Research Collaboration Fund of the University of Nottingham.

\appendix

\section{Explicit expression of the optical Bloch equation} \label{A}

The dynamics of the atomic motion is governed by the optical Bloch equation
\begin{align}
\label{Bloch0}
\frac{\partial \hat{\rho}}{\partial t}=-\frac{i}{\hbar}\left[\hat{H}, \hat{\rho}\right]-\Gamma\,[\hat{\rho}].
\end{align}
Here $\hat{\rho}$ is the density matrix (DM) describing the atomic population and coherence, with the DM elements defined by $\rho_{\alpha\beta}\equiv\langle\hat{S}_{\alpha\beta}\rangle$; $\Gamma$ is the relaxation matrix describing the spontaneous emission and dephasing.

Based on the Hamiltonian $\hat{H}$ given in the main text, we obtain the explicit expression of the optical Bloch equation with the following form
\begin{subequations}\label{Bloch1}
\begin{align}
& i \frac{\partial}{\partial t} \rho_{11}-i \Gamma_{13} \rho_{33}+i \Gamma_{21} \rho_{11}+\Omega_p^* \rho_{31}-\Omega_p \rho_{13}=0 \\
& i \frac{\partial }{\partial t}\rho_{22}-i \Gamma_{23} \rho_{33}-i \Gamma_{21} \rho_{11}+\Omega_c^* \rho_{32}-\Omega_c \rho_{23}=0, \\
& i \frac{\partial}{\partial t}\rho_{33}-i \Gamma_{34} \rho_{44}+i \Gamma_3 \rho_{33}-\Omega_p^* \rho_{31}+\Omega_p \rho_{13}-\Omega_c^* \rho_{32}\notag\\
&~\qquad+\Omega_c \rho_{23}+\Omega_d^* \rho_{43}-\Omega_d \rho_{34}=0, \\
& i \frac{\partial}{\partial t} \rho_{44}+i \Gamma_{34} \rho_{44}-\Omega_d^* \rho_{43}+\Omega_d \rho_{34}=0,
\end{align}
\end{subequations}
for the diagonal elements, and
\begin{subequations}\label{Bloch2}
\begin{align}
& \left(i \frac{\partial}{\partial t}+d_{21}\right) \rho_{21}+\Omega_c^* \rho_{31}-\Omega_p \rho_{23}=0, \\
& \left(i \frac{\partial}{\partial t}+d_{31}\right) \rho_{31}+\Omega_d^* \rho_{41}+\Omega_p\left(\rho_{11}-\rho_{33}\right)+\Omega_c \rho_{21}=0, \\
& \left(i \frac{\partial}{\partial t}+d_{32}\right) \rho_{32}+\Omega_d^* \rho_{42}+\Omega_p \rho_{12}+\Omega_c\left(\rho_{22}-\rho_{33}\right)=0, \\
& \left(i \frac{\partial}{\partial t}+d_{41}\right) \rho_{41}+\Omega_d \rho_{31}-\Omega_p \rho_{43}\notag\\
&~\qquad-\mathcal{N}_a \int d^3 \mathbf{r}^{\prime} V\left(\mathbf{r}^{\prime}-\mathbf{r}\right) \rho_{44,41}\left(\mathbf{r}^{\prime}, \mathbf{r}, t\right)=0, \\
& \left(i \frac{\partial}{\partial t}+d_{42}\right) \rho_{42}+\Omega_d \rho_{32}-\Omega_c \rho_{43}\notag\\
&~\qquad-\mathcal{N}_a \int d^3 \mathbf{r}^{\prime} V\left(\mathbf{r}^{\prime}-\mathbf{r}\right) \rho_{44,42}\left(\mathbf{r}^{\prime}, \mathbf{r}, t\right)=0, \\
& \left(i \frac{\partial}{\partial t}+d_{43}\right) \rho_{43}-\Omega_p^* \rho_{41}-\Omega_c^* \rho_{42}+\Omega_d\left(\rho_{33}-\rho_{44}\right)\notag\\
&~\qquad-\mathcal{N}_a \int d^3 \mathbf{r}^{\prime} V\left(\mathbf{r}^{\prime}-\mathbf{r}\right) \rho_{44,43}\left(\mathbf{r}^{\prime}, \mathbf{r}, t\right)=0,
\end{align}
\end{subequations}for the non-diagonal elements.
Here $d_{\alpha\beta}=\Delta_{\alpha}-\Delta_{\beta}+i \gamma_{\alpha\beta}$, $\gamma_{\alpha\beta}=(\Gamma_{\alpha}+\Gamma_{\beta})/2$, $\Gamma_{\alpha}=\sum_{\alpha<\beta}\Gamma_{\alpha\beta}$, with $\Gamma_{\alpha\beta}$ the spontaneous emission decay rate from $|\beta\rangle$ to $|\alpha\rangle$; $\Gamma_{21}$ is the rate of population exchange between $|1\rangle$ and $|2\rangle$; $\rho_{44,4\alpha}({\bf r^{\prime},r},t)=\langle \hat{S}_{44}({\bf r^{\prime}},t) \hat{S}_{4\alpha}({\bf r},t)\rangle$ are two-body
DM elements; the interaction between two Rydberg
atoms respectively at positions $\mathbf{r}$ and $\mathbf{r}^{\prime}$ is described by the potential
$V\left(\mathbf{r}^{\prime}-\mathbf{r}\right)=-
\hbar C_{6} / | \mathbf{r}^{\prime}-\mathbf{r}| ^6$, with $C_6$ the dispersion parameter.

\section{Solution of the MB equations} \label{B}

\subsection{Solutions of one-body density-matrix elements}

Since the probe field is much weaker than the control and dressed fields, we can take $\Omega_{p}$ 
as an expansion parameter and the perturbation expansion
$\rho_{\alpha\alpha}=\rho_{\alpha\alpha}^{(0)}+\varepsilon
\rho_{\alpha\alpha}^{(1)}+\varepsilon ^2\rho_{\alpha\alpha}^{(2)}+\cdots$ ($\alpha=1,\,2,\,3,\,4$), and
$\rho_{\alpha\beta}=\varepsilon \rho_{\alpha\beta}^{(1)}+\varepsilon ^2\rho_{\alpha\beta}^{(2)}+\cdots$\, ($\alpha=2,\,3,\,4;\beta=1,\,2,\,3;\alpha>\beta$). Substituting the above expansions into the Eqs.~(\ref{Bloch1}) and (\ref{Bloch2}), we obtain a set of linear but inhomogeneous equations which can be solved order by order~\cite{newell1992book}.

At the zeroth order, the solutions read
\begin{subequations}
\begin{align}
\rho_{11}^{(0)}= & 2 \Gamma_{13}\left(2 Z^2+2 X Y+X \Gamma_{34}\right) / M, \\
\rho_{22}^{(0)}= & \Gamma_{21}\big[4\left(Z^2+X Y\right)+2(X+Z) \Gamma_{34}\notag\\
&-\left(\Gamma_{13}+\Gamma_{23}\right)\left(2 Y+\Gamma_{34}\right)\big] / M, \\
\rho_{33}^{(0)}= & 2 \Gamma_{21}\left[2\left(Z^2+X Y\right)+X \Gamma_{34}\right] / M, \\
\rho_{44}^{(0)}= & 2 \Gamma_{21}\left[2\left(Z^2+X Y\right)-Z\left(\Gamma_{13}+\Gamma_{23}\right)\right] / M, \\
\rho_{32}^{(0)}= & {\left[(|\Omega_c|^2-d_{42} d_{43}) \rho_{22}^{(0)} -\left|\Omega_d\right|^2 \rho_{44}^{(0)}\right.}\notag \\
& \left. +(|\Omega_d|^2-\left|\Omega_c\right|^2+d_{42} d_{43}) \rho_{33}^{(0)}\right] \Omega_c / D_1, \\
\rho_{42}^{(0)}= & {\left[d_{43} \rho_{22}^{(0)}-\left(d_{32}+d_{43}\right) \rho_{33}^{(0)}+d_{32} \rho_{44}^{(0)}\right] \Omega_c \Omega_d / D_1, } \\
\rho_{43}^{(0)}= & \big[\left|\Omega_c\right|^2 \rho_{22}^{(0)}+(\left|\Omega_d\right|^2-\left|\Omega_c\right|^2+d_{42} d_{32}) \rho_{33}^{(0)}\notag\\
&\qquad-(\left|\Omega_d\right|^2-d_{42} d_{32}) \rho_{44}^{(0)}\big] / D_1,
\end{align}
\end{subequations}
where $M=\Gamma_{21}[12\left(X Y+Z^2\right)+2 \Gamma_{34}(2 X+Z)]+\Gamma_{13}\left[2 \Gamma_{34} X+4 X Y-\Gamma_{21}\left(\Gamma_{34}+2 Y+2 Z\right)+4 Z^2\right]-\Gamma_{21}\Gamma_{23}\left[\Gamma_{34}+2(Y+Z)\right]$, 
$X=2\rm{Im}[(d_{42} d_{43}-\left|\Omega_c\right|^2) / D_1]\left|\Omega_c\right|^2$, 
$Y=2\rm{Im}[(\left|\Omega_d\right|^2-d_{42} d_{32}) /D_1]\left|\Omega_d\right|^2$, 
$Z=2\rm{Im}\left[1 / D_1\right]|\Omega_d|^2|\Omega_c|^2$, 
and $D_1=d_{32} d_{42} d_{43}-d_{32}\left|\Omega_c\right|^2-d_{43}\left|\Omega_d\right|^2$.

From the zeroth order solution, we find that
the incoherent population pumping rate $\Gamma_{21}$ is a key parameter in the zeroth order solution. If $\Gamma_{21}=0$, all populations are in the ground state, i.e., $\rho_{11}^{(0)}=1$, and other state population and coherence are both zero, $\rho_{\alpha\beta}^{(0)}=0$.
However, when $\Gamma_{21}\ne0$, we have $\rho_{33}^{(0)}\ne0$, and hence a gain to the probe ﬁeld will be realized when the probe ﬁeld is coupled to the states $|1\rangle$ and $|3\rangle$.

{\flushleft{\bf First-order solutions}}:
At the first order, 
the solutions of $\rho_{21}^{(1)}$, $\rho_{31}^{(1)}$, and $\rho_{41}^{(1)}$ are given by
\begin{subequations}
\begin{align}
\rho_{21}^{(1)}&= \Big[\left(d_{31} d_{41}-\left|\Omega_d\right|^2\right) \rho_{32}^{*(0)}-d_{41} \Omega_c\left(\rho_{33}^{(0)}-\rho_{11}^{(0)}\right)\notag\\
&\qquad+\Omega_d \Omega_c \rho_{43}^{(0)}\Big] / D_2 \Omega_p \equiv a_{21}^{(1)} \Omega_p, \\
\rho_{31}^{(1)}&= \Big[-\Omega_c d_{41} \rho_{32}^{*(0)}+d_{21} d_{41}\left(\rho_{33}^{(0)}-\rho_{11}^{(0)}\right)\notag\\
&\qquad-d_{21} \Omega_d \rho_{43}^{(0)}\Big] / D_2 \Omega_p \equiv a_{31}^{(1)} \Omega_p,\\
\rho_{41}^{(1)} &=  \Big[\Omega_c \Omega_d \rho_{32}^{*(0)}-d_{21} \Omega_d\left(\rho_{33}^{(0)}-\rho_{11}^{(0)}\right)\notag\\
&\qquad+\left(d_{21} d_{31}-\left|\Omega_c\right|^2\right) \rho_{43}^{(0)}\Big] / D_2 \Omega_p 
\equiv a_{41}^{(1)} \Omega_p,
\end{align}
\end{subequations}
where $D_2=d_{31}|\Omega_c|^2+d_{21}|\Omega_d|^2-d_{21}d_{31}d_{41}$. Other $\rho_{\alpha \beta}^{(1)}$ are zero.

{\flushleft{\bf Second-order solutions}}: At the second order, the matrix elements can be solved by the equation
\begin{footnotesize}
\begin{align}
&\left(\begin{array}{cccccccccc}
i \Gamma_{21} & 0 & -i \Gamma_{13} & 0 & 0 & 0 & 0 & 0 & 0 & 0 \\
-i \Gamma_{21} & 0 & -i \Gamma_{23} & 0 & 0 & 0 & 0 & 0 & \Omega_c^* & -\Omega_c \\
0 & 0 & 0 & i \Gamma_{34} & -\Omega_d^* & \Omega_d & 0 & 0 & 0 & 0 \\
0 & \Omega_c & -\Omega_c & 0 & 0 & 0 & \Omega_d^* & 0 & d_{32} & 0 \\
0 & 0 & 0 & 0 & -\Omega_c & 0 & d_{42} & 0 & \Omega_d & 0 \\
0 & 0 & \Omega_d & -\Omega_d & d_{43} & 0 & -\Omega_c^* & 0 & 0 & 0 \\
0 & \Omega_c^* & -\Omega_c^* & 0 & 0 & 0 & 0 & \Omega_d & 0 & d_{32}^* \\
0 & 0 & 0 & 0 & 0 & -\Omega_c^* & 0 & d_{42}^* & 0 & \Omega_d^* \\
0 & 0 & \Omega_d^* & -\Omega_d^* & 0 & d_{43}^* & 0 & -\Omega_c & 0 & 0 \\
1 & 1 &
1 & 1 & 0 & 0 & 0 & 0 & 0 & 0
\end{array}\right)\notag\\
&~\qquad
\left(\begin{array}{c}
\rho_{11}^{(2)} \\
\rho_{22}^{(2)} \\
\rho_{33}^{(2)} \\
\rho_{44}^{(2)} \\
\rho_{43}^{(2)} \\
\rho_{34}^{(2)} \\
\rho_{42}^{(2)} \\
\rho_{24}^{(2)} \\
\rho_{32}^{(2)} \\
\rho_{23}^{(2)}
\end{array}\right)=\left(\begin{array}{c}
2 i {\rm Im}\left[\Omega_p \rho_{13}^{(1)}\right] \\
0 \\
0 \\
-\Omega_p \rho_{12}^{(1)} \\
0 \\
\Omega_p^* \rho_{41}^{(1)} \\
-\Omega_p^* \rho_{21}^{(1)} \\
0 \\
\Omega_p \rho_{14}^{(1)} \\
0
\end{array}\right) . 
\end{align}
\end{footnotesize}
Solving the matrix equations above yields $\rho_{\alpha\beta}^{(2)}=a_{\alpha\beta}^{(2)}|\Omega_p|^2$ with the coefficients $a_{\alpha\beta}$ being the function detuning $\Delta_{\alpha}$, spontaneous emission decay rate  $\Gamma_{\alpha\beta}$, and half Rabi frequencies $\Omega_d$, $\Omega_c$.

{\flushleft{\bf Third-order solutions}}: At the third order, the solutions of $\rho_{21}^{(3)}$, $\rho_{31}^{(3)}$, and $\rho_{41}^{(3)}$ can be obtain from the equations
\begin{align}\label{third-order}
\left(\begin{array}{ccc}
d_{21} & \Omega_c^* & 0 \\
\Omega_c & d_{31} & \Omega_d^* \\
0 & \Omega_d & d_{41}
\end{array}\right)\left(\begin{array}{c}
\rho_{21}^{(3)} \\
\rho_{31}^{(3)} \\
\rho_{41}^{(3)}
\end{array}\right)=\left(\begin{array}{c}
a_{23}^{(2)} \\
a_{33}^{(3)}-a_{11}^{(2)} \\
a_{43}^{(2)}
\end{array}\right)\left|\Omega_p\right|^2 \Omega_p\notag\\+
\left(\begin{array}{ccc}
0 &
0 &
\mathbb{A}
\end{array}\right)^T,
\end{align}
where $\mathbb{A}=\mathcal{N}_a \int d{\bf r}' V({\bf r'}-{\bf r})a_{44,41}|\Omega_p({\bf r'})|^2\Omega_p$. Expressions of $\rho_{31}^{(3)}$ at the third order is obtained from Eq.~(\ref{third-order})
\begin{align}
\rho_{31}^{(3)}=a_{31}^{(3)}\left|\Omega_p\right|^2 \Omega_p+\mathcal{N}_a \int d^3 \mathbf{r}^{\prime} V\left(\mathbf{r}^{\prime}-\mathbf{r}\right) b_{31}^{(3)}\left|\Omega_p\left(\mathbf{r}^{\prime}\right)\right|^2 \Omega_p,
\end{align}
where $a_{31}^{(3)}=[d_{21} d_{41}(a_{33}^{(2)}-a_{11}^{(2)})-d_{41} \Omega_c a_{32}^{*(2)}-d_{21} \Omega_d^* a_{43}^{(2)}]$ $ / D_2$ and $b_{31}^{(3)}=$ $d_{21} \Omega_d^* a_{44,41}^{(3)}\left(\mathbf{r}^{\prime}-\mathbf{r}\right) / D_2$. Combining first three order solutions of $\rho_{31}$ with Maxwell equation Eq.~(3), we obtain the nonlocal nonlinear schr\"odinger equation
\begin{align}
&i \frac{\partial \Omega_p}{\partial z}+\frac{c}{2 \omega_p} \nabla_{\perp}^2 \Omega_p-V_1 \Omega_p+W\left|\Omega_p\right|^2 \Omega_p
\notag\\&\qquad
+\int d \mathbf{r}^3 G\left(\mathbf{r}^{\prime},\mathbf{r}\right)\left|\Omega_p\left(\mathbf{r}^{\prime}\right)\right|^2 \Omega_p=0,
\end{align}
where $V_1=-\kappa_{13} a_{31}^{(1)}$ is linear potential, $W=\kappa_{13}[d_{21} d_{41}(a_{33}^{(2)}-a_{11}^{(2)})-d_{41} \Omega_c a_{32}^{*(2)}-d_{21} \Omega_d^* a_{43}^{(2)}]/ D_2$ is the local nonlinear coefficient, and the nonlocal response function $G\left(\mathbf{r}^{\prime},\mathbf{r}\right)=\left(\kappa_{13} d_{21} \Omega_d^* \mathcal{N}_a / D_2\right) V\left(\mathbf{r}^{\prime}-\mathbf{r}\right)$ $ a_{44,41}^{(3)}\left(\mathbf{r}^{\prime}-\mathbf{r}\right)$. Note that the two-body equations $a_{44,41}^{(3)}$ should be solved firstly in order to solve the nonlocal nonlinear Schr\"odinger equation above. 

\begin{figure}[t]
\centering
\includegraphics[width=0.55\linewidth]{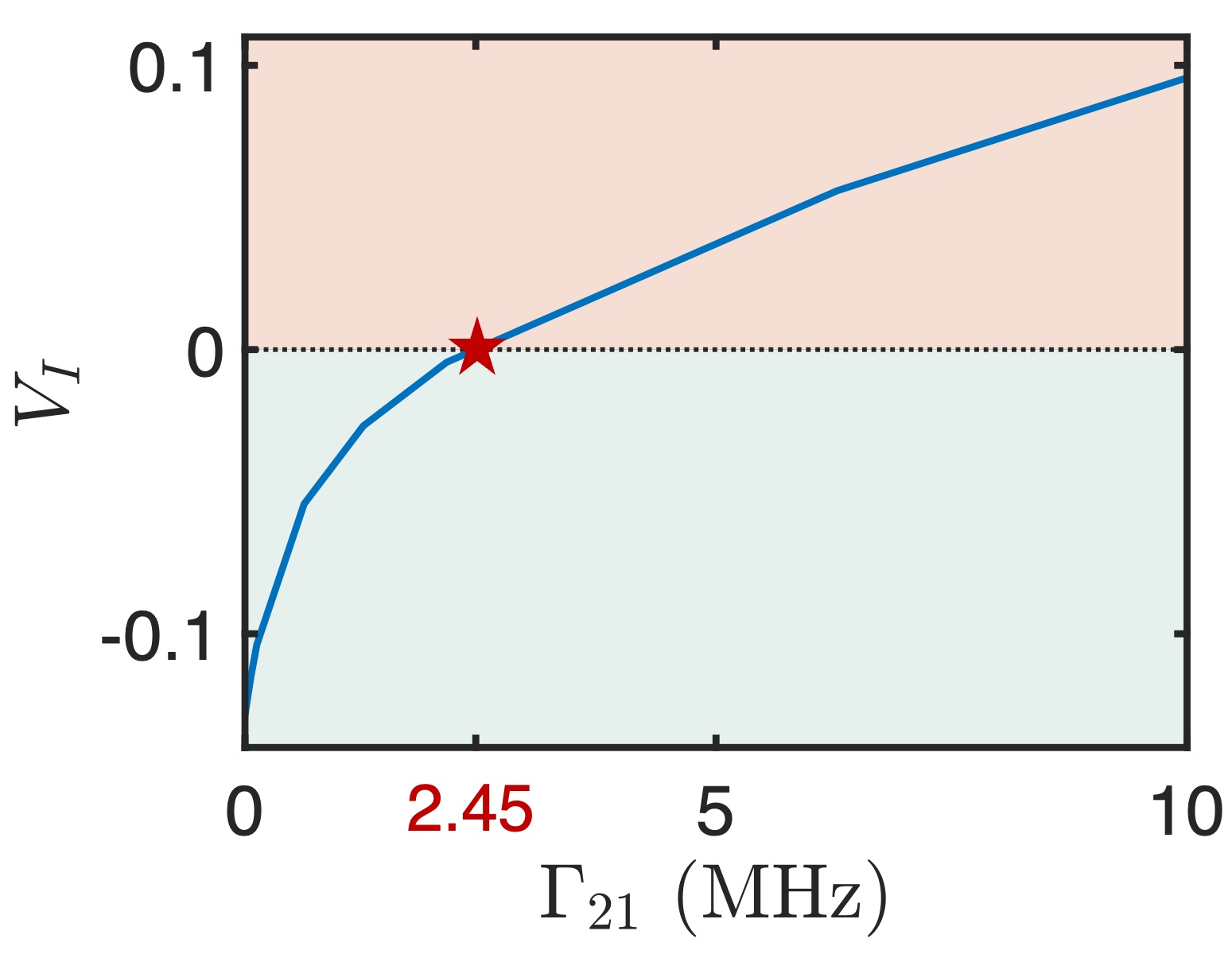}
\caption{(a)~The imaginary potential as function as incoherent pumping $\Gamma_{21}$. 
When $\Gamma_{21}<2.45$~MHz, the potential is a loss. Otherwise, the potential is a gain. The red star represents $V_I=0$ when $\Gamma_{21}=2.45$~MHz. The lower green and upper orange regions mark the loss and gain potential, respectively. 
}\label{fig_imaginary_potential}
\end{figure}

In dimensionless NNLS equation Eq.~(\ref{DE1}), the linear potential reads
\begin{align}
\mathcal{V}=&-2\kappa_{13}L_{\rm diff}\Big[-\Omega_c d_{41} \rho_{32}^{*(0)}+d_{21} d_{41}\left(\rho_{33}^{(0)}-\rho_{11}^{(0)}\right)\notag\\
&\qquad-d_{21} \Omega_d \rho_{43}^{(0)}\Big] / D_2 \equiv V_R+iV_I. 
\end{align}
Here, $V_R$ and $V_I$ are the real and imaginary parts of linear potential, respectively. $V_R$ is a constant and independent on the intensity of the shock wave, and is indeed negligible. $V_I<0$ show a loss to the probe ﬁeld due to the dissipation of system when 
$\rho_{33}^{(0)}=0$ in the zeroth order for $\Gamma_{21}=0$. With increasing of incoherent pumping $\Gamma_{21}$, $\rho_{33}^{(0)}\ne0$ and a gain to the probe ﬁeld will be realized. As results show in Fig.~\ref{fig_imaginary_potential}, a loss/gain potential $V_I$ can be realized by adjusting the $\Gamma_{21}$. When $\Gamma_{21}<2.45$~MHz, the potential is a loss. Otherwise, the potential is a gain.

\subsection{Solutions of two-body density-matrix elements}

We solve the second order solution of the correlators of $\rho _{41,41}^{(2)}$ which can be obtained from
\begin{align}\small
&\left( {\begin{array}{*{20}{c}}
2d_{41} - V&{2{\Omega_d}}&0&0&0&0\\
{\Omega_d^*}&{{d_{41}} + {d_{31}}}&{{\Omega_d}}&{{\Omega _c}}&0&0\\
0&{\Omega_d^*}&{{d_{31}}}&0&{{\Omega _c}}&0\\
0&{\Omega _c^*}&0&{{d_{41}} + {d_{21}}}&{{\Omega_d}}&0\\
0&0&{\Omega _c^*}&{\Omega_d^*}&{{d_{31}} + {d_{21}}}&{{\Omega _c}}\\
0&0&0&0&{\Omega _c^*}&{{d_{21}}}
\end{array}} \right)
 \notag\\&~\qquad
\left( {\begin{array}{*{20}{c}}
{\rho _{41,41}^{(2)}}\\
{\rho _{41,31}^{(2)}}\\
{\rho _{31,31}^{(2)}}\\
{\rho _{41,21}^{(2)}}\\
{ \rho _{31,21}^{(2)}}\\
{ \rho _{21,21}^{(2)}}
\end{array}} \right)
= \left( {\begin{array}{*{20}{c}}
2a_{43}^{(0)}a_{41}^{(1)}\\
(a_{33}^{(0)}-a_{11}^{(0)})a_{41}^{(1)}+a_{43}^{(0)}a_{31}^{(1)}\\
(a_{33}^{(0)}-a_{11}^{(0)})a_{31}^{(1)}\\
a_{32}^{*(0)}a_{41}^{(1)}+a_{43}^{(0)}a_{21}^{(1)}\\
(a_{33}^{(0)}-a_{11}^{(0)})a_{21}^{(1)}+a_{23}^{(0)}a_{31}^{(1)}\\
a_{32}^{*(0)}a_{21}^{(1)}
\end{array}} \right){\Omega_p^2}.
\end{align}

The solution for  ${\rho _{41,41}^{(2)}}=a_{41,41}^{(2)}({\bf r'-r})\Omega_p^2({\bf r'})$  with
\begin{align}
  a_{41,41}^{(2)}({\bf r'-r})&=\frac{P_0}{P_1+P_2V({\bf r'-r})},
\end{align}
where  $P_0$, $P_1$, and $P_2$ are the functions of $\Omega_d$, $\Omega_c$, $\Delta_{\alpha}$, and $\Gamma_{\alpha\beta}$.

The  two-body  equations for $\rho_{44,41}^{(3)}$ are 27 order linear equations, which are very lengthy and hence are omitted here, and  solution has the form
\begin{align}
\rho_{44,41 }^{(3)}({\bf{r'}} - {\bf{r}})& = \frac{\sum_{n=0}^2P_nV^n({\bf r'}-{\bf r})}{\sum_{n=0}^3Q_nV^ n({\bf r'}-{\bf r})}|\Omega_p({\bf r'})|^2\Omega_p({\bf r})\notag\\
&\approx \rho_{41,41}^{(2)}\rho_{14}^{(1)},
\end{align}
where $P_n$, $Q_n$ are the functions of $\Omega_d$, $\Omega_c$, $\Delta_{\alpha}$, and $\Gamma_{\alpha\beta}$.

\begin{figure}[t]
\centering
\includegraphics[width=1\linewidth]{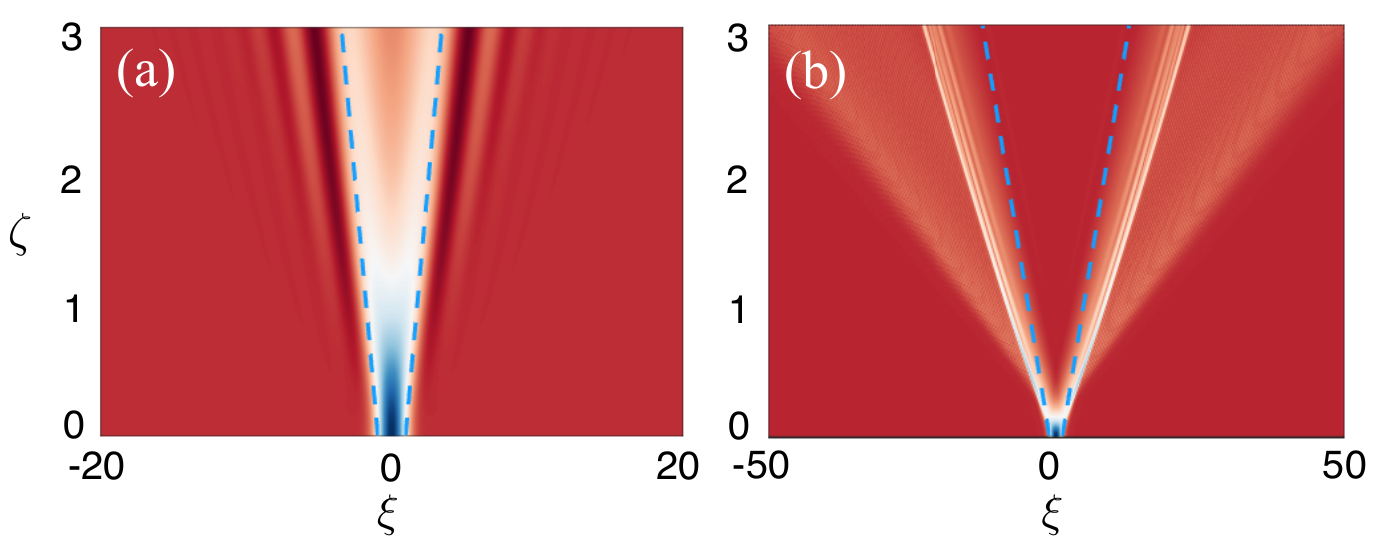}
\caption{\footnotesize
(a)~Wave propagation when $\sigma=0.5$. The dashed line marks the trajectory of the sound wave. It is clear that the wave propagates at the speed of sound $c_s$, indicating the system is in the linear regime. Other parameters are same as in Fig.~\ref{fig:propagation1}(a), corresponding to the linear propagation discussed in Sec.~\ref{sec:nonlocalA}. (b) The same as (a). Other parameters are same as in Fig.~\ref{fig:propagation2}(a), corresponding to the linear propagation discussed in Sec.~\ref{sec:nonlocalB}. 
}\label{fig_propagation_withcs}
\end{figure}

Note that the response function in dimensionless NNLS equation Eq.~(\ref{DE1}) can be obtain from the two-body correlators $\rho_{44,41}^{(3)}$. The expression reads
$g=2L_{\rm diff}U_0^2R_0^2 \kappa_{13} d_{21} \Omega_d^* \mathcal{N}_a / D_2 \int V\left(\mathbf{r}^{\prime}-\mathbf{r}\right)$ $ a_{44,41}^{(3)}dy'$.
It can be approximated by analytical form~\cite{hang2018PRA,Hang2023PRA}
\begin{align}\label{response function1}
g(\xi',\xi)&\approx- \int \frac{1}{b_1+ \frac{b_2}{\sigma^6} \left[(\xi'-\xi)^2+(y'/R_0)^2 \right]^3}dy'\notag\\
&\approx -\frac{B_1}{B_2\sigma^6+|\xi'-\xi|^6},
\end{align}
where $\sigma=R_b/R_0$ characterizes the nonlocality degree of the  nonlinearity. 
$b_1$ and $b_2$ are the coefficients determined by laser parameters (i.e. $\Omega_d$, $\Omega_c$, $\Delta_{\alpha}$, and $\Gamma_{\alpha\beta}$). The relation between $B_{1,2}$ and $b_{1,2}$ are $B_1=\sigma^6/b_2$ and $B_2=b_1/b_2$.

 \newpage
\nocite{*}

\bibliography{sample}

\end{document}